\documentclass[format=acmsmall, review=false, screen]{acmart}

\makeatletter
\def\@ACM@checkaffil{%
    \if@ACM@instpresent\else
    \ClassWarningNoLine{\@classname}{No institution present for an affiliation}%
    \fi
    \if@ACM@citypresent\else
    \ClassWarningNoLine{\@classname}{No city present for an affiliation}%
    \fi
    \if@ACM@countrypresent\else
        \ClassWarningNoLine{\@classname}{No country present for an affiliation}%
    \fi
}
\makeatother

\acmConference{}{}{} %
\startPage{0} %
\AtBeginDocument{
\fancypagestyle{firstpagestyle}{%
	\fancyhf{}%
	\fancyfoot[L]{\footnotesize{}}%
}
\fancypagestyle{ourstyle}{%
	\fancyhf{}%
	\fancyhead[C]{\footnotesize{\mbox{\shortauthors}}}
	\fancyhead[R]{\thepage}%
}
\pagestyle{ourstyle}
\renewcommand{\footnotetextcopyrightpermission}[1]{} %
\renewcommand{\footnotetextauthorsaddresses}[1]{} %
}

\usepackage{booktabs} %
\usepackage[ruled]{algorithm2e} %

\SetAlFnt{\small}
\SetAlCapFnt{\small}
\SetAlCapNameFnt{\small}
\SetAlCapHSkip{0pt}
\IncMargin{-\parindent}
\allowdisplaybreaks

\setcitestyle{authoryear}

\title{Beyond the Limits of Welfarism: An Affordable Proportional Approval Voting Rule}

\author{Anton Baychkov}

\begin{abstract}
    Two of the most widely studied voting rules in approval-based multiwinner elections are Proportional Approval Voting (PAV), which maximizes a welfare score over committees of a fixed size, and the Method of Equal Shares (MES), which gives voters equal budgets and selects candidates whose supporters can pay for them. MES guarantees priceability and laminar proportionality, both of which can fail for PAV. Can PAV offer similar guarantees without giving voters budgets? To answer this question, we go beyond the limits of welfarism by allowing PAV to select committees of varying size, while penalizing larger committees. Both our exact rule and its polynomial-time local-search variant produce outcomes that admit a priceable completion and, on laminar instances, a laminar-proportional one. Our analysis carefully separates the roles of two marginal stability conditions. At suitable thresholds, addition stability gives a strong proportionality guarantee in the form of EJR+, while deletion stability limits overrepresentation and ensures affordability. We use these conditions to derive welfare and coverage guarantees for different completions of our rules. The same welfare comparison also improves the best known lower bound on the utilitarian ratio of standard PAV.
    \end{abstract}

\usepackage{nicefrac}
\usepackage{tikz}
\usepackage{tikz-cd}
\usepackage{cleveref}
\usepackage{thmtools, thm-restate} %
\usepackage{enumitem}
\usepackage{wrapfig}
\usepackage{wrapstuff}
\usepackage{arydshln} %
\usepackage{graphicx}
\usepackage{tabularx}
\usepackage{colortbl} %

\newif\ifincludetightness
\includetightnessfalse

\makeatletter
\renewcommand{\@parfont}{\normalfont\bfseries}
\makeatother

\theoremstyle{acmplain}
\newtheorem{theorem}{Theorem}
\newtheorem{lemma}[theorem]{Lemma}
\newtheorem{corollary}[theorem]{Corollary}
\newtheorem{proposition}[theorem]{Proposition}
\newtheorem{observation}{Observation}

\declaretheorem[
  style=acmdefinition,
  title=Example,
  refname={example,examples},
  Refname={Example,Examples},
  qed={$\vartriangle$}
]{example}

\declaretheorem[
  style=acmdefinition,
  title=Definition,
  refname={definition,definitions},
  Refname={Definition,Definitions},
  qed={$\vartriangle$}
]{definition}

\newcommand{\indicator}{\mathbf{1}}

\newcommand{\score}{\mathrm{score}}

\usepackage{xcolor}
\usepackage{pifont}
\newcommand{\cmark}{\textcolor{green!70!black}{\ding{51}}}  %

\newcommand{\anton}[1]{{\color{olive}{AB: #1 }}}
\newcommand{\mb}[1]{{\color{teal}{\ifcomments MB: #1 \fi}}}

\begin{document}

\begin{titlepage}

\maketitle

\vspace{1cm}
\setcounter{tocdepth}{1} %
\tableofcontents

\end{titlepage}

\section{Introduction}

Multiwinner elections involve selecting a committee of $k$ candidates based on the preferences of $n$ voters. We consider approval-based ballots, with each voter indicating the subset of candidates they approve of \citep{LaSk22a}. Two of the most prominent proportional rules for such elections are Proportional Approval Voting (PAV) \citep{Thie95a} and the Method of Equal Shares (MES) \citep{PeSk20b}. Both satisfy strong proportionality axioms such as Extended Justified Representation+ (EJR+) \citep{BrPe23a}, but they arrive at their outcomes in different ways. PAV selects a committee of $k$ candidates that maximizes a score based on voter satisfaction. MES gives voters equal budgets and selects candidates whose supporters can pay for them. In \emph{Proportionality and the Limits of Welfarism}, \citet{PeSk20b} show that MES satisfies the two normative properties of priceability and laminar proportionality. Neither of these can be guaranteed by a Pareto-optimal welfarist rule, such as PAV.

One of the hidden benefits of MES is that it can output a partial outcome that leaves some seats unfilled. This outcome is affordable -- if each voter has a budget of $\nicefrac{k}{n}$, then the supporters of each selected candidate can together pay one unit for that candidate without any voter exceeding their budget. The payment system constructed by MES can therefore help explain the outcome to voters and stakeholders \citep{PPSS21a}. If all $k$ seats must be filled, the remaining seats can be used to pursue welfare or coverage guarantees \citep{BrPe24a}. Alternatively, the outcome can be pruned to avoid overrepresentation, while keeping its EJR+ guarantee intact \citep{BLMNP26}\footnote{The authors argue this for GJCR, but their argument can easily be extended to MES.}. 
Since PAV always optimizes for a committee of exactly $k$ candidates, it is natural to ask whether allowing it to leave seats unfilled could give it some of the same flexibility.

We study a simple modification of the PAV objective that allows the committee size to vary, but introduces a penalty that is a fixed multiple of the committee size. This penalty
determines how much a candidate must contribute to the PAV score to justify an additional seat. Our exact rule, Exact Penalized PAV (EP-PAV), sets the multiplier to the Hare quota $\nicefrac{n}{k}$ \citep{ABC+16a}. We also investigate a polynomial-time local-search variant, Local-Search Penalized PAV (LP-PAV). Both rules return a committee satisfying affordability, in addition to EJR+. Thus a payment system exists, even though neither rule utilizes voter budgets to select its committee. We investigate a selection of further properties satisfied by our rules, including axioms concerning underrepresentation and overrepresentation, as well as properties of various completions to committees of size $k$.

With our results, we revisit the dichotomy between MES and PAV established by \citet{PeSk20b}. Their impossibility results concern welfarist rules, which maximize some function of the vector of voter satisfaction among committees of size $k$. Our exact rule retains the
PAV score but compares committees using both voter satisfaction and committee size. We find that the resulting penalized PAV (P-PAV) rules produce outcomes that admit priceable and laminar-proportional completions. MES and P-PAV nevertheless
offer differing proportionality guarantees, with MES ensuring Strong EJR+ \citep{Skow26b} and P-PAV retaining the $(\ell-1)$-representativeness of PAV.

\noindentparagraph{Our Contributions.}
We separate the roles of addition stability, which places an upper bound on the PAV score gained by adding any unselected candidate, and deletion stability, which places a lower bound on the PAV score lost by removing any selected candidate. For an appropriately chosen bound, addition stability gives EJR+ (\Cref{prop:PAV_improvement_EJR}), while deletion stability guarantees affordability (\Cref{prop:PAV_deletion_affordability}), upper quota (\Cref{prop:PAV_decline_upper_quota}), and justified near quota (\Cref{prop:PAV_deletion_JNQ}). We identify thresholds at which both conditions hold for our two rules LP-PAV and EP-PAV (\Cref{cor:PAV_polynomial_guarantees,cor:PAV_exact_rule}). For EP-PAV, we also give an explicit payment system in which each winner's supporters share its cost in proportion to their contributions to the marginal PAV score (\Cref{prop:PAV_simple_payments}). Combining the two stability conditions, we prove a utilitarian guarantee of $\nicefrac{2}{\sqrt{k}}-\nicefrac{2}{k}$ for Approval Voting (AV) completion (\Cref{prop:PAV_utilitarian_completion}). The same welfare comparison yields an improved utilitarian guarantee for standard PAV and its polynomial-time local-search variant
(\Cref{cor:regular_PAV_utilitarian}). For Chamberlin--Courant (CC) completion, affordability lets us apply the results of \citet{BrPe24a} to obtain a coverage guarantee of $\nicefrac{3}{4}$ (\Cref{cor:PAV_CC_completion}).

\section{Related Work}

\noindentparagraph{Thiele rules, PAV and local search.}
Thiele rules, encompassing AV, PAV, CC, and many others, provide a common framework for selecting a committee based on voter welfare. PAV is singled out as the only rule satisfying EJR within this family \citep{ABC+16a}, while its restriction to apportionment instances yields the D'Hondt apportionment method \citep{BLS18a}. For fixed-size committees, PAV's EJR guarantee can also be obtained in polynomial time through a bounded local search based on candidate swaps \citep{SLES17a,AEH+18a}. This bounded approach is important, as unrestricted local search can require super-polynomially many steps \citep{KrEl24a}.
More recently, \citet{PVY26a} analyze the Pareto-efficiency guarantees of local PAV and its approximation of the optimal PAV score, while \citet{CaEl26a} establish Droop-based proportionality guarantees, including Droop-EJR+ for a polynomial-time local-search PAV rule.

\noindentparagraph{Penalized objectives and other optimization approaches.}

Our size-penalized PAV objective has the same form as the Lagrangian relaxation used by \citet{MaSo26a} to compute exact PAV-maximizing committees on voter-interval profiles. They use the penalty to recover a committee of size $k$, while we fix it at $\nicefrac{n}{k}$ when defining EP-PAV and study the resulting partial outcome. The bounded PAV improvement condition studied by \citet{GoYa26a} requires that adding any unselected candidate increases the PAV score by strictly less than $\nicefrac{n}{k}$. They connect this condition to Lindahl priceability in apportionment and show that it implies EJR+ and $2$-frugal Lindahl priceability in committee elections. Their arguments for these two guarantees also apply to partial committees. We investigate what further guarantees follow from imposing a corresponding condition on deletions.

Related optimization approaches extend beyond the ordinary PAV score. \citet{BGP26a} optimize a harmonic entropy function over committees and payment systems to ensure Core+, a property stronger than EJR+. Similarly, \citet{Skow26a} combines utility-weighted harmonic entropy with a linear cost penalty to obtain Core-up-to-one in participatory budgeting. In the multiwinner setting, the resulting rule uses the same penalty and tie-breaking as EP-PAV, with utility-weighted harmonic entropy in place of the PAV score. In concurrent work (see \hyperref[sec:acks]{Acknowledgments and AI Use}), \citet{AzVo26a} modify the harmonic entropy rule in a similar way, showing in particular that Core+ is compatible with affordability, and additionally consider a local-search variant that is computable in polynomial time.

\noindentparagraph{Partial outcomes and completion.}
When introducing MES, \citet{PeSk20b} noted that it may leave seats unfilled and proposed completing its outcome using, for instance, sequential Phragm\'en. The notion of completion also connects representation to the literature on welfare and coverage \citep{LaSk20b,EFI+22a}. \citet{BrPe24a} develop a framework for completing affordable partial committees with strong guarantees on these objectives.
Affordable partial outcomes also play a role in best-of-both-worlds committee voting \citep{ALS+23a}.
In particular, \citet{SuVo24a} show that every affordable committee admits a polynomial-time computable
lottery over its completions satisfying group resource proportionality ex ante, providing an additional connection between affordability and proportional representation.
Finally, \citet{BLMNP26} show that allowing seats to remain unfilled makes it possible to combine standard proportionality guarantees with axioms limiting overrepresentation. We show that their notions of upper and near quota can follow from deletion stability.

\section{Preliminaries}\label{sec:preliminaries}

\newcommand{\bound}{{\ensuremath\lambda}}
\newcommand{\pav}{\score_{PAV}}
\newcommand{\apav}{\score_{PAV-\bound}}

\noindentparagraph{Approval-based multiwinner voting.}
A multiwinner voting instance consists of a finite candidate set $C$,
a voter set $N=\{1,\ldots,n\}$, an approval profile
$(A_i)_{i\in N}$ with $A_i\subseteq C$, and a target
committee size $k$, with $n\geq1$ and $1\leq k\leq|C|$.
For $c\in C$, let $N(c)=\{i\in N:c\in A_i\}$ denote its supporter set.
We assume that every candidate has at least one
supporter, that is, $N(c)\neq\emptyset$ for every $c\in C$.
We say that voter $i$'s satisfaction from $W\subseteq C$ is
$|A_i\cap W|$. A committee $W\subseteq C$ is \emph{feasible}
if $|W|\leq k$, and \emph{exhaustive} if $|W|=k$.
A \emph{completion} of a feasible committee $W$ is a size-$k$
superset of $W$. A feasible committee $W$ is \emph{Pareto-optimal} if there is
no feasible committee $W'$ with $|A_i\cap W'|\geq |A_i\cap W|$
for every $i\in N$ and strict inequality for at least one voter.

We use the Hare quota $n/k$ for all quota-dependent proportionality notions.
The \emph{quota} of a voter group $N'\subseteq N$ is defined to be $q(N')=\frac{k|N'|}{n}$, and the quota of a particular candidate $c$ is $q_c=q(N(c))$.
The \emph{lower} and \emph{upper quotas} of $N'$ are $\lfloor q(N')\rfloor$
and $\lceil q(N')\rceil$, respectively.
Note that these quotas depend on the target size $k$, even when $|W|\neq k$.

\noindentparagraph{Thiele voting rules.}
A \emph{weight function}
$w:\mathbb{N}_{>0}\rightarrow\mathbb{R}_{\geq0}$ satisfies
$w(j)\geq w(j+1)$ for every $j\geq1$ and is normalized by $w(1)=1$.
For any outcome $W$ we denote its $w$-\emph{Thiele} score by:
$$\score_w(W)
=\sum_{i\in N}
\sum_{j=1}^{|A_i\cap W|}w(j).$$

The $w$-Thiele rule selects all size-$k$ committees maximizing this score. Three notable Thiele rules are \emph{Proportional Approval Voting} (PAV), with $w(j)=1/j$, \emph{Approval Voting} (AV), with $w(j)=1$, and \emph{Chamberlin-Courant} (CC), with $w(1)=1$ and $w(j)=0$ for all $j\geq 2$.

AV maximizes \emph{utilitarian welfare} over size $k$ committees, $\operatorname{sw}(W)=\sum_{i\in N}|A_i\cap W|=\sum_{c\in W}|N(c)|$. We define the \emph{utilitarian ratio} of $W$ as $\operatorname{sw}(W)/\operatorname{sw}(W^*)$, where $W^*$ is an AV-winning committee. A utilitarian guarantee is a lower bound on this ratio over all relevant instances and outcomes \citep{LaSk20b,BrPe24a}. Similarly, CC maximizes \emph{coverage}, $\operatorname{cov}(W)=|\{i\in N:A_i\cap W\neq\emptyset\}|$. We define the \emph{representation ratio} of $W$ as $\operatorname{cov}(W)/\operatorname{cov}(W^*)$, where $W^*$ is a coverage-maximizing committee and the \emph{representation guarantee} as the lower bound on this ratio.

For $|W|\leq k$, its \emph{AV completion} adds the $k-|W|$ candidates with the most supporters outside $W$, and its \emph{CC completion} is a size-$k$ superset maximizing coverage.
A \emph{sequential CC completion} of $W$ repeatedly adds a candidate covering the most previously unrepresented voters until the committee has size $k$.

\noindentparagraph{Proportionality axioms.}

\begin{definition}\label{def:EJR_plus}
An outcome $W$ satisfies \emph{extended justified representation plus}
(EJR+) \cite{BrPe23a} if there is no $c\in C\setminus W$ and
nonempty $N'\subseteq N(c)$ such that for every $i\in N'$
\begin{equation*}
    |A_i\cap W|<\lfloor q(N')\rfloor.\qedhere
\end{equation*}
\end{definition}

We use \emph{Strong EJR+} as shorthand for the approval,
unit-cost specialization of \emph{Strong EJR+ up to one}
from \citet[Definition~4]{Skow26b}.

\begin{definition}\label{def:strong_EJR_plus}
An outcome $W$ satisfies \emph{Strong EJR+} if, for every
$c\in C\setminus W$ and every $N'\subseteq N$ with
$N'\cap N(c)\neq\emptyset$, there is a voter $i\in N'$ such that
\begin{equation*}
\sum_{c'\in(W\cup\{c\})\cap A_i}
\frac{1}{
\max\{|N'\cap N(c')|,|N'\cap N(c)|\}}
>\frac{k}{n}.\qedhere
\end{equation*}
\end{definition}

\begin{definition}\label{def:quota_axioms}
An outcome $W$ satisfies \emph{upper quota} (UQ) \citep{BLMNP26} if there is no
$c\in W$ such that
$$
|A_i\cap W|>\lceil q_c\rceil
\qquad\text{for every }i\in N(c).
$$
It satisfies \emph{justified upper quota} (JUQ) if there is no
$c\in W$, $d\in C\setminus W$, and nonempty $N'\subseteq N(d)$
such that
\begin{align*}
|A_i\cap W|
&>\lceil q_c\rceil
&&\text{for every }i\in N(c),\\
|A_i\cap((W\setminus\{c\})\cup\{d\})|
&\leq\lceil q(N')\rceil
&&\text{for every }i\in N'.
\end{align*}
It satisfies \emph{justified near quota} (JNQ) if there is no
$c\in W$, $d\in C\setminus W$, and nonempty $N'\subseteq N(d)$
such that
\begin{align*}
|A_i\cap W|-q_c
&>\frac12\indicator[d\notin A_i]
&&\text{for every }i\in N(c),\\
q(N')-|A_i\cap W|
&>\frac12\indicator[c\notin A_i]
&&\text{for every }i\in N'.\qedhere
\end{align*}
\end{definition}

Here, the indicator $\indicator[P]$ equals $1$ if $P$ holds and $0$ otherwise.

For a function $f:\{1,\ldots,k\}\rightarrow\mathbb{R}$, a set $W$
is \emph{$f(\ell)$-representative} \cite{BrPe23a} if, for every
integer $1\leq\ell\leq k$, every $c\notin W$, and every
$N'\subseteq N(c)$ with $q(N')\geq\ell$,
$$
\frac1{|N'|}\sum_{i\in N'}|A_i\cap W|\geq f(\ell).
$$

\noindentparagraph{Payment-based axioms.}
Candidates have unit cost, and a total budget $B$ gives each
voter $B/n$.

\begin{definition}\label{def:affordability}
For $B>0$, a set $W$ is \emph{affordable with budget $B$} if
there are voter payment functions
$(p_i:C\rightarrow[0,\nicefrac{B}{n}])_{i\in N}$ satisfying:
\begin{enumerate}
\item $c\notin A_i\Rightarrow p_i(c)=0$ for all $i\in N$
      and $c\in C$;
\item $\sum_{c\in C}p_i(c)\leq B/n$ for all $i\in N$;
\item $\sum_{i\in N}p_i(c)=1$ for all $c\in W$;
\item $\sum_{i\in N}p_i(c)=0$ for all $c\notin W$.\qedhere
\end{enumerate}
\end{definition}
We call $W$ \emph{affordable} if it is affordable with budget $k$ \cite{BrPe24a}. A set $W$ is \emph{priceable} \cite{PeSk20b} if, for some $B>0$,
the four conditions above hold together with

\begin{enumerate}
 \setcounter{enumi}{4}
\item $\sum_{i\in N(c)}
\left(\frac Bn-\sum_{d\in C}p_i(d)\right)\leq1$ for all $c\notin W$.
\end{enumerate}

For Lindahl priceability, we use the same normalization as affordability: each voter has budget $k/n$ and each candidate has unit cost. Personalized prices describe hypothetical purchases, including unselected candidates, rather than actual payments.

\begin{definition}\label{def:frugal_Lindahl}
For $\alpha\geq1$, a set $W$ is \emph{$\alpha$-frugal Lindahl priceable} \cite{GoYa26a} if there are nonnegative personalized prices $p_i(c)$ such that:
\begin{enumerate}
\item $\sum_{i\in N}p_i(c)\leq 1$
      for every $c\in C$;
\item for every $i\in N$ and $C'\subseteq C$,
      $\sum_{c\in C'}p_i(c)\leq{k/n}$ implies
      $|A_i\cap C'|\leq\alpha|A_i\cap W|$;
\item $p_i(c)\leq p_i(d)$ whenever
      $c\in A_i\cap W$ and $d\in A_i\setminus W$.\qedhere
\end{enumerate}
\end{definition}

For feasible committees, $1$-frugal Lindahl priceability is equivalent to \emph{Core+} \cite[Proposition~A.4]{BGP26a}.

\noindentparagraph{The Method of Equal Shares.}
The \emph{Method of Equal Shares} (MES) \cite{PeSk20b} starts with an empty committee $W=\emptyset$ and initial voter budgets $b_i=k/n$. Among all unselected candidates $c\notin W$ with
$\sum_{i\in N(c)}\min\{b_i,\rho\}=1,$ for some $\rho\geq 0$, it chooses a $c$ with minimum $\rho$, adds $c$ to $W$, and charges each supporter $i$ the amount $\min\{b_i,\rho\}$. It terminates when no such candidates exist.

\noindentparagraph{Restricted instance classes.}
In an \emph{apportionment instance}, voters are partitioned into nonempty parties $N_1,\ldots,N_t$ with pairwise disjoint candidate lists $C_1,\ldots,C_t$, where $|C_j|=k$ and $A_i=C_j$ for every $i\in N_j$ \citep{BLS18a}. A committee satisfies \emph{lower quota} if $|W\cap C_j|\geq\lfloor q(N_j)\rfloor$ for every $j$, and \emph{exact lower quota} if equality holds for every $j$.

A multiwinner voting instance is \emph{laminar} if one of the following recursive clauses applies. Each clause also specifies its \emph{laminar-proportional} committees \citep[Definitions~2 and~3]{PeSk20b}.
\begin{enumerate}
\item If $A_i=C$ for every $i\in N$ and $|C|\geq k$,
      then every size-$k$ subset of $C$ is laminar proportional.
\item Suppose $c$ is approved by every voter. If deleting $c$
      from $C$ and all ballots yields a nonunanimous laminar
      instance at target size $k-1$, then the laminar-proportional
      committees are exactly the sets $W'\cup\{c\}$, where $W'$
      is laminar proportional in the reduced instance.
\item If $N=N_1\sqcup N_2$ and $C=C_1\sqcup C_2$, with
      $A_i\subseteq C_j$ for every $i\in N_j$, and the two
      subinstances on $N_j$ and $C_j$ are laminar at target
      sizes $k_j$ satisfying $k_1+k_2=k$ and
      $\nicefrac{|N_1|}{k_1}=\nicefrac{|N_2|}{k_2}$,
      then the laminar-proportional committees are exactly
      the unions $W_1\cup W_2$, where each $W_j$ is laminar
      proportional in its subinstance.
\end{enumerate}

\section{The Penalized PAV Rules}\label{sec:affordable_PAV}

In this section we consider a version of the usual PAV objective, penalized by a \bound-multiple of the committee size, and use it to define an exact penalized PAV rule. We consider two corresponding notions of local stability, $\bound$-addition stability and $\bound$-deletion stability, and use them to devise a local-search variant. We then investigate which familiar axioms are implied by these stability notions and for which values of $\bound$ these relationships hold.

\medskip

For $\bound\geq 0$ we define the \emph{\bound-penalized PAV objective} to be
$$\apav(W)=\pav(W)-\bound|W|.$$

We can use this objective for the following voting rule.

\begin{definition}\label{def:ED_PAV}
    \emph{Exact Penalized PAV} (EP-PAV), parameterized by some penalty $\bound\geq 0$, outputs all maximum-cardinality committees from among those maximizing $\apav$ over all $W\subseteq C$.
\end{definition}

EP-PAV retains a close connection to the welfarist rules studied by \citet{PeSk20b}. For a given target size $k$, the only information its objective uses, aside from the vector of voter satisfaction, is the size of the committee. In particular, if EP-PAV returns a committee of size $s$, that committee maximizes the ordinary PAV score among all committees of size $s$. The penalty only matters when comparing committees with different sizes. For the guarantees established below, our canonical choice is $\bound=\nicefrac{n}{k}$ (\Cref{cor:PAV_exact_rule}).

To define a local search variant of our penalized rule, we define the following stability conditions. 

\begin{definition}\label{def:PAV_stability}
For $\bound\geq0$, a set $W\subseteq C$ is
\emph{$\bound$-addition-stable for PAV} if, for every $c\in C\setminus W$,
$$
\pav(W\cup\{c\})-\pav(W)\leq\bound,
$$
and \emph{$\bound$-deletion-stable for PAV} if, for every $c\in W$,
$$
\pav(W)-\pav(W\setminus\{c\})\geq\bound.
$$
It is \emph{$\bound$-stable for PAV} if both conditions hold.
\end{definition}

Thus $\bound$-stability is precisely local optimality of $\apav$ under single-candidate additions and deletions. Unlike local optimality for the standard PAV rule, we do not fix the committee size or require stability under swaps.

\begin{definition}\label{def:LD_PAV}
    \emph{Local-Search Penalized PAV} (LP-PAV) is parametrized by an addition threshold $\bound_1$ and a deletion threshold $\bound_2$, with $\bound_1>\bound_2\geq 0$. Starting from $W=\emptyset$, repeatedly perform any applicable move:
    \begin{enumerate}
        \item Add $c\notin W$ if $\pav(W\cup\{c\})-\pav(W)>\bound_1$.
        \item Delete $c\in W$ if $\pav(W)-\pav(W\setminus\{c\})<\bound_2$.
        \item Replace $c\in W$ by $d\notin W$ if $N(c)\subsetneq N(d)$.
    \end{enumerate}
    When no move is applicable, return $W$.
\end{definition}

At termination, LP-PAV finds a committee that is $\bound_1$-addition-stable and $\bound_2$-deletion-stable, as a consequence of conditions (1) and (2) respectively. The gap between the thresholds allows us to bound the number of local-search moves. In particular, we canonically choose $\bound_1=\frac nk-\frac{1}{3k(k+1)}$ and $\bound_2=\frac nk-\frac{2}{3k(k+1)}$ to obtain a polynomial-time rule (\Cref{prop:PAV_local_search}). For this choice of thresholds, $\bound_1$-addition-stability and $\bound_2$-deletion-stability imply many of the axioms considered further in this section (\Cref{cor:PAV_polynomial_guarantees}). Finally, condition~(3) ensures that no selected candidate has a supporter set strictly contained in that of an unselected candidate, which allows us to find a laminar proportional completion (\Cref{prop:PAV_laminar_completion}).

\medskip

The following example distinguishes MES, PAV, and Penalized PAV.

\begin{example}
Consider the approval profile in \Cref{tab:EP_PAV_example}, with $n=11$ and $k=3$. 

Every PAV or EP-PAV winner contains $c_1$. Indeed, replacing any of $c_2,c_3,c_4$ by $c_1$ strictly increases the PAV score, since $N(c_1)$ strictly contains each of their supporter sets. This leaves only subsets of $\{c_5\}$, which have fewer than $k$ members and for which adding $c_1$ gains $9>\bound=\nicefrac{11}{3}$. The same swap or addition is available to LP-PAV, so every terminal committee it outputs must also contain $c_1$.

\begin{table}[h!]
    \centering
    \begin{tabular}{c|cccc}
        Candidate & $A_1$--$A_5$ & $A_6$--$A_7$
                  & $A_8$--$A_9$ & $A_{10}$--$A_{11}$ \\
        \hline
        $c_1$ & \cmark & \cmark & \cmark &        \\
        $c_2$ & \cmark & \cmark &        &        \\
        $c_3$ &        & \cmark & \cmark &        \\
        $c_4$ & \cmark &        &        &        \\
        $c_5$ &        &        &        & \cmark \\
    \end{tabular}
    \caption{Approval profile with $k=3$.}
    \label{tab:EP_PAV_example}
\end{table}
    The unique EP-PAV outcome is $\{c_1\}$. Once $c_1$ is selected, adding another candidate gains at most $\nicefrac{7}{2}$, attained by $c_2$. Since this is below $\bound$, any further additions decrease the penalized score. It is also below the canonical deletion threshold $\bound_2$, so any other selected candidate is eligible for deletion. Thus LP-PAV also returns $\{c_1\}$, regardless of the order of moves.

    The unique MES outcome is $\{c_1,c_2\}$. The most popular candidate $c_1$ is selected first, leaving voters $1,\ldots,9$ with $\nicefrac{16}{99}$ each. The supporters of $c_2,c_3,c_4,c_5$ have total remaining budgets of $\nicefrac{112}{99}$, $\nicefrac{64}{99}$, $\nicefrac{80}{99}$, and $\nicefrac{6}{11}$, respectively. Only $c_2$ is affordable, so MES selects it and stops.

    The unique PAV outcome is $\{c_1,c_2,c_5\}$, as can be checked by comparing the six possible pairs of candidates that can accompany $c_1$. This outcome is not priceable. Only voters $10$ and $11$ can fund $c_5$, so each voter's budget must be at least $\nicefrac{1}{2}$. Voters $1,\ldots,9$ collectively spend $2$ on $c_1$ and $c_2$,
    leaving them at least $\nicefrac{5}{2}$ in total. Since $N(c_3)\cup N(c_4)=\{1,\ldots,9\}$, the supporters of at least one of $c_3$ and $c_4$ have more than one unit of budget remaining, contradicting priceability.\qedhere
\end{example}

\medskip

\begin{figure}[h!]
\centering
\begingroup
\usetikzlibrary{arrows.meta}

\definecolor{PAVBlue}{HTML}{2166C5}
\definecolor{PAVRed}{HTML}{C43B43}
\definecolor{PAVInk}{HTML}{20232A}
\definecolor{PAVMuted}{HTML}{626873}
\definecolor{PAVGrid}{HTML}{E6E8ED}
\definecolor{PAVPurple}{HTML}{7653A5}

\def\xLabel{0}
\def\xZero{6.1}
\def\xSize{7.45}            %
\def\xAffordable{9.25}      %
\def\xQuota{11.05}          %
\def\xRepresentative{12.9}  %
\def\xEnd{13.6}
\def\xReference{14.7}
\def\yTopAxis{2.1}
\def\yAxis{24.4}
\def\lineStep{1.4}    %
\def\noteStep{1.0}    %
\def\textCenter{.34}  %
\def\yLegend{27.5}
\def\legendStep{1.4}
\def\xLegendEndpoint{5}

\def\yRulesTop{\yTopAxis}
\def\yRulesBottom{\yAxis}
\def\ldSideInset{1/3} %

\tikzset{
  pavfig/line/.style={line width=.85pt},
  pavfig/rule dots/.style={
    line width=.85pt,dash pattern=on \pgflinewidth off 2.5pt
  },
  pavfig/lp fill/.style={fill=PAVPurple,fill opacity=.12,draw=none},
  pavfig/label/.style={anchor=west},
  pavfig/note/.style={anchor=west,font=\small,text=PAVMuted},
  pavfig/reference/.style={anchor=west,font=\small},
  pavfig/point/.style={
    circle,anchor=center,inner sep=0pt,outer sep=0pt,
    minimum size=4.5pt,line width=.6pt
  },
  pavfig/guide/.style={draw=PAVGrid,line width=.3pt}
}

\def\PAVClosed{closed}
\def\PAVTriangle{triangle}
\def\PAVRay{ray}
\newcommand{\PAVMarker}[3]{%
  \def\PAVKind{#3}%
  \ifx\PAVKind\PAVTriangle
    \begin{scope}[shift={#2},x=1pt,y=1pt]
      \draw[draw=#1,fill=#1,line width=.6pt]
        (0,2.6) -- (-2.3,-1.3) -- (2.3,-1.3) -- cycle;
    \end{scope}
  \else
    \ifx\PAVKind\PAVClosed
      \node[pavfig/point,draw=#1,fill=#1] at #2 {};
    \else
      \node[pavfig/point,draw=#1,fill=white] at #2 {};
    \fi
  \fi
}
\newcommand{\PAVRange}[6]{%
  \def\PAVKind{#6}%
  \ifx\PAVKind\PAVRay
    \draw[pavfig/line,draw=#1,-{Stealth[length=4.5pt]}]
      (#3,{#2}) -- (#5,{#2});
  \else
    \draw[pavfig/line,draw=#1] (#3,{#2}) -- (#5,{#2});
    \PAVMarker{#1}{(#5,{#2})}{#6}
  \fi
  \PAVMarker{#1}{(#3,{#2})}{#4}
}
\newcommand{\PAVLabel}[3]{%
  \node[pavfig/label,anchor=base west] at (\xLabel,#1) {#2};
  \if\relax\detokenize{#3}\relax
  \else
    \node[pavfig/note,anchor=base west]
      at (\xLabel,{#1+\noteStep}) {#3};
  \fi
}
\newcommand{\PAVRow}[4]{%
  \PAVLabel{#1}{#2}{#3}
  \if\relax\detokenize{#3}\relax
    \node[pavfig/reference] at (\xReference,{#1-\textCenter}) {\Cref{#4}};
  \else
    \node[pavfig/reference]
      at (\xReference,{#1+.5*\noteStep-\textCenter}) {\Cref{#4}};
  \fi
}
\newcommand{\PAVPair}[2]{%
  \draw[draw=PAVMuted,line width=.5pt]
    ({\xZero-.1},{#1-.25}) -- ({\xZero-.25},{#1-.25})
    -- ({\xZero-.25},{#2+.25}) -- ({\xZero-.1},{#2+.25});
  \node[anchor=east,font=\small,text=PAVMuted]
    at ({\xZero-.42},{(#1+#2)/2}) {Both};
  \draw[draw=PAVMuted,line width=.5pt]
    ({\xEnd+.1},{#1-.25}) -- ({\xEnd+.25},{#1-.25})
    -- ({\xEnd+.25},{#2+.25}) -- ({\xEnd+.1},{#2+.25});
}
\newcommand{\PAVLPBand}[4]{%
  \path[pavfig/lp fill] (#1,{#3}) rectangle (#2,{#4});
  \draw[pavfig/rule dots,draw=PAVRed] (#1,{#3}) -- (#1,{#4});
  \draw[pavfig/rule dots,draw=PAVBlue] (#2,{#3}) -- (#2,{#4});
}
\newcommand{\PAVTickLabel}[4]{%
  \node[anchor=#3] at (#1,#2) {$\vphantom{\frac{1}{k(k+1)}}#4$};
}
\newcommand{\PAVAxis}[3]{%
  \draw[-{Stealth[length=4pt]},thin] (\xZero,#1) -- ({\xEnd+.12},#1);
  \foreach \x in {\xZero,\xSize,\xAffordable,\xQuota,\xRepresentative}
    \draw[thin] (\x,{#1-.14}) -- (\x,{#1+.14});
  \node[anchor=west,font=\Large] at ({\xEnd+.25},#1) {$\bound$};
  \PAVTickLabel{\xZero}{#2}{#3}{0}
  \PAVTickLabel{\xSize}{#2}{#3}{\frac{n}{k+1}}
  \PAVTickLabel{\xAffordable}{#2}{#3}{\frac{n}{k}-\frac{1}{k(k+1)}}
  \PAVTickLabel{\xQuota}{#2}{#3}{\frac{n}{k}}
  \PAVTickLabel{\xRepresentative}{#2}{#3}{\frac{n}{k}+\frac{1}{k(k-1)}}
}

\resizebox{\linewidth}{!}{%
\begin{tikzpicture}[
  x=1cm,y=-.42cm,font=\normalsize,text=PAVInk,line cap=round,
  every node/.style={inner sep=0pt,outer sep=0pt}
]

\node[pavfig/label,font=\bfseries] at (\xLabel,\yTopAxis) {Guarantee};
\node[pavfig/reference,font=\bfseries] at (\xReference,\yTopAxis) {Result};

\foreach \x in {\xZero,\xSize,\xAffordable,\xQuota,\xRepresentative}
  \draw[pavfig/guide] (\x,\yTopAxis) -- (\x,\yAxis);
\pgfmathsetmacro{\xLPDeletion}{\xAffordable+\ldSideInset*(\xQuota-\xAffordable)}
\pgfmathsetmacro{\xLPAddition}{\xQuota-\ldSideInset*(\xQuota-\xAffordable)}

\PAVLPBand{\xLPDeletion}{\xLPAddition}{\yRulesTop}{\yRulesBottom}
\draw[pavfig/rule dots,draw=PAVInk]
  (\xQuota,{\yRulesTop}) -- (\xQuota,{\yRulesBottom});

\PAVAxis{\yTopAxis}{1.45}{south}

\PAVRow{3.8}{EJR+}{}{prop:PAV_improvement_EJR}
\PAVRange{PAVBlue}{3.8-\textCenter}{\xZero}{closed}{\xQuota}{triangle}

\PAVRow{5.2}{$(\ell-1)$-representativeness}{}{prop:PAV_improvement_representativeness}
\PAVRange{PAVBlue}{5.2-\textCenter}{\xZero}{closed}{\xRepresentative}{open}

\PAVRow{6.6}{Committee size $|W|\le k$}{}{prop:PAV_decline_size}
\PAVRange{PAVRed}{6.6-\textCenter}{\xSize}{open}{\xEnd}{ray}

\PAVRow{8}{Upper quota (UQ)}{}{prop:PAV_decline_upper_quota}
\PAVRange{PAVRed}{8-\textCenter}{\xSize}{open}{\xEnd}{ray}

\PAVRow{9.4}{Justified upper quota (JUQ)}{}{prop:PAV_deletion_JUQ}
\PAVRange{PAVRed}{9.4-\textCenter}{\xSize}{closed}{\xEnd}{ray}

\PAVRow{10.8}{Justified near quota (JNQ)}{}{prop:PAV_deletion_JNQ}
\PAVRange{PAVRed}{10.8-\textCenter}{\xAffordable}{closed}{\xEnd}{ray}

\PAVRow{12.2}{Affordability}{}{prop:PAV_deletion_affordability}
\PAVRange{PAVRed}{12.2-\textCenter}{\xAffordable}{open}{\xEnd}{ray}

\PAVRow{13.6}{Simple affordable payments}{}{prop:PAV_simple_payments}
\PAVRange{PAVRed}{13.6-\textCenter}{\xQuota}{closed}{\xEnd}{ray}

\PAVRow{15}{Representation guarantee \nicefrac{3}{4}}
  {with CC completion}{cor:PAV_CC_completion}
\PAVRange{PAVRed}{15+.5*\noteStep-\textCenter}{\xAffordable}{open}{\xEnd}{ray}

\PAVRow{17.4}{Laminar proportionality\textsuperscript{$\dagger$}}
  {with MES completion}{prop:PAV_laminar_completion}
\PAVRange{PAVRed}{17.4+.5*\noteStep-\textCenter}{\xSize}{open}{\xEnd}{ray}

\PAVRow{19.8}{Utilitarian guarantee $\nicefrac{2}{\sqrt{k}}-\nicefrac{2}{k}$}
  {with AV completion}{prop:PAV_utilitarian_completion}
\PAVPair{19.8-\textCenter}{19.8+\noteStep-\textCenter}
\PAVRange{PAVBlue}{19.8-\textCenter}{\xZero}{closed}{\xQuota}{closed}
\PAVRange{PAVRed}{19.8+\noteStep-\textCenter}{\xAffordable}{open}{\xEnd}{ray}

\PAVRow{22.2}{Exact lower quota}
  {on apportionment instances}{prop:exact_lower_quota}
\PAVPair{22.2-\textCenter}{22.2+\noteStep-\textCenter}
\PAVRange{PAVBlue}{22.2-\textCenter}{\xZero}{closed}{\xQuota}{triangle}
\PAVRange{PAVRed}{22.2+\noteStep-\textCenter}{\xAffordable}{open}{\xEnd}{ray}

\PAVAxis{\yAxis}{25.05}{north}

\draw[pavfig/line,PAVBlue] (0,\yLegend) -- (.35,\yLegend);
\node[pavfig/note,text=PAVInk] at (.5,\yLegend)
  {$\bound$-addition-stability};
\draw[pavfig/line,PAVRed]
  (0,{\yLegend+\legendStep}) -- (.35,{\yLegend+\legendStep});
\node[pavfig/note,text=PAVInk]
  at (.5,{\yLegend+\legendStep}) {$\bound$-deletion-stability};

\draw[pavfig/rule dots,draw=PAVInk]
  (.175,{\yLegend+2*\legendStep-.38})
  -- (.175,{\yLegend+2*\legendStep+.38});
\node[pavfig/note,text=PAVInk]
  at (.5,{\yLegend+2*\legendStep}) {EP-PAV};
\PAVLPBand{.06}{.29}
  {\yLegend+3*\legendStep-.38}{\yLegend+3*\legendStep+.38}
\node[pavfig/note,text=PAVInk]
  at (.5,{\yLegend+3*\legendStep}) {LP-PAV};

\PAVMarker{PAVInk}{(\xLegendEndpoint,\yLegend)}{open}
\node[pavfig/note,text=PAVInk] at ({\xLegendEndpoint+.22},\yLegend)
  {Strict bound};
\PAVMarker{PAVInk}{(\xLegendEndpoint,{\yLegend+\legendStep})}{closed}
\node[pavfig/note,text=PAVInk]
  at ({\xLegendEndpoint+.22},{\yLegend+\legendStep}) {Weak bound};
\PAVMarker{PAVInk}{(\xLegendEndpoint,{\yLegend+2*\legendStep})}{triangle}
\node[pavfig/note,text=PAVInk]
  at ({\xLegendEndpoint+.22},{\yLegend+2*\legendStep})
  {Endpoint included for optimization with maximum-cardinality tiebreaking};

\end{tikzpicture}%
}
\endgroup
\caption{Properties implied by \bound-addition stability and \bound-deletion stability.
Blue (red) ranges give sufficient values of $\bound$ for which \bound-addition (deletion) stability implies that property. Properties satisfied by the two penalized PAV rules for our canonical choice of thresholds are indicated; if the dotted line (EP-PAV) or shaded area (LP-PAV) intersects a particular range, then our rule satisfies the corresponding property.
\\The scale is schematic and is shown for $k\geq2$.
\\$\dagger$~The laminar completion guarantee assumes that there
are no $c\in W$ and $d\notin W$ with $N(c)\subsetneq N(d)$.
}
\label{fig:PAV_stability_ranges}
\end{figure}
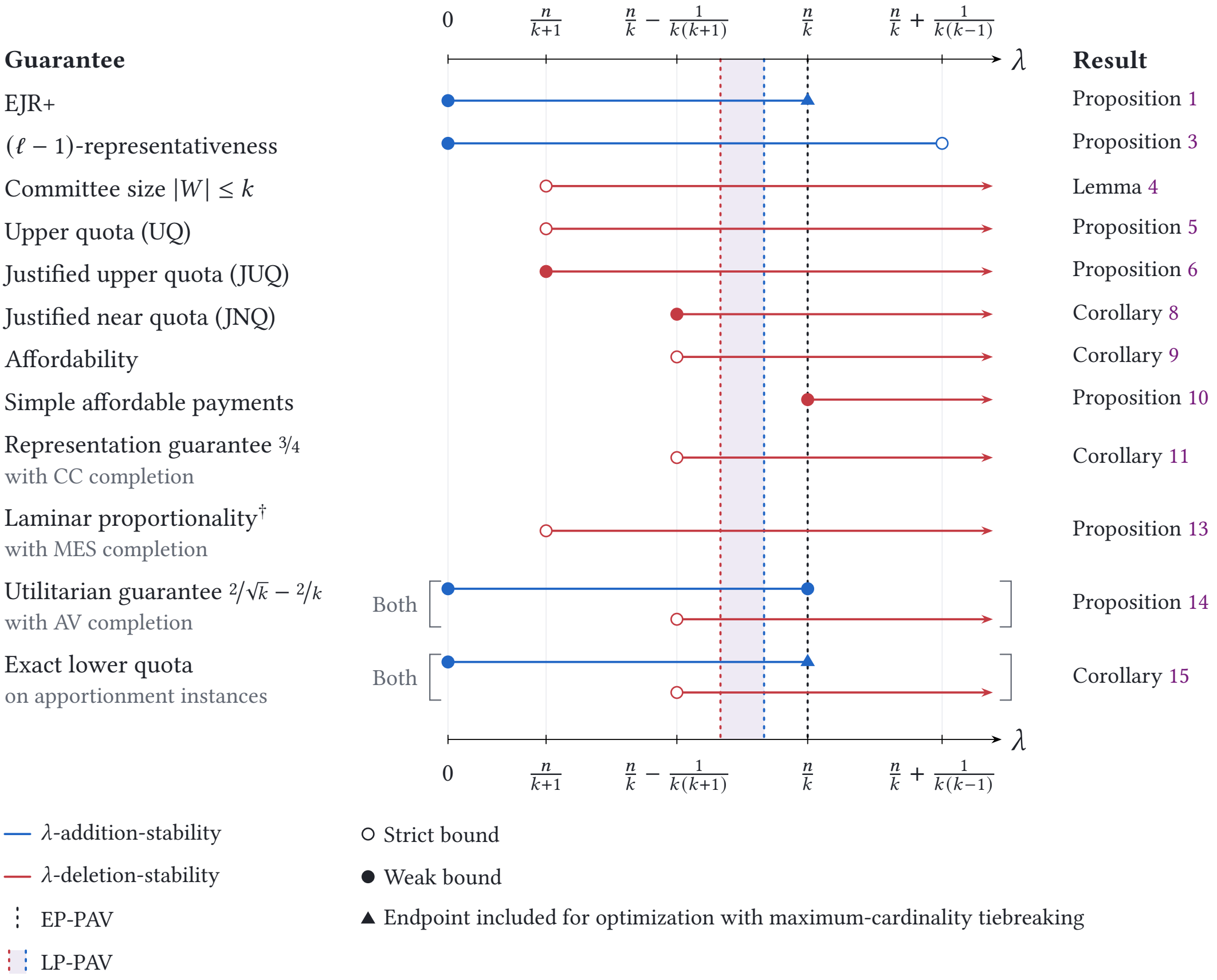

For the next three subsections, we investigate the properties implied by $\bound$-addition and $\bound$-deletion stability for different values of $\bound$. Observe that increasing $\bound$ makes addition stability a weaker property, as this allows candidates with a higher potential contribution to the marginal PAV score to be excluded. Similarly, increasing $\bound$ makes deletion stability stronger. \Cref{fig:PAV_stability_ranges} summarizes the guarantees obtained from each stability condition. 

\subsection{Consequences of Addition Stability}\label{sec:PAV_addition}

As a warm-up, we consider some consequences of addition stability, showing that many of the proportionality guarantees of regular PAV carry over to the penalized objective. These results build on the marginal PAV arguments of \citet{ABC+16a} and their extension to EJR+ and representativeness by \citet{BrPe23a}. Following \citet{GoYa26a} we say that $W$ satisfies \emph{bounded PAV improvement} if
$$
\pav(W\cup\{c\})-\pav(W)<\frac nk
\qquad\text{for every }c\in C\setminus W.
$$

Our notion of addition stability generalizes this to other values of $\bound$.

\begin{observation}\label{obs:bounded_PAV_improvement}
If $W$ is $\bound$-addition-stable with
$\bound<\nicefrac{n}{k}$, then $W$ satisfies bounded PAV improvement.
\end{observation}

Via \Cref{obs:bounded_PAV_improvement}, the next two implications follow directly from \citet[Proposition~D.1 and Theorem~4.2]{GoYa26a}, respectively, by observing that their proofs also apply to partial committees. The representativeness result below sharpens the argument of their Proposition~D.3.

\begin{proposition}\label{prop:PAV_improvement_EJR}
If $W$ is $\bound$-addition-stable
with $\bound<\nicefrac{n}{k}$, then $W$ satisfies EJR+.
\end{proposition}

\begin{proposition}\label{prop:PAV_improvement_frugal_Lindahl}
If $W$ is $\bound$-addition-stable
with $\bound<\nicefrac{n}{k}$, then $W$ is $2$-frugal Lindahl
priceable.
\end{proposition}

\begin{proposition}\label{prop:PAV_improvement_representativeness}
If $W$ is $\bound$-addition-stable
with $\bound<\frac{n}{k}+\frac{1}{k(k-1)}$ and $k\geq 2$, then $W$ is
$(\ell-1)$-representative.
\end{proposition}

Note that, for $k=1$, $(\ell-1)$-representativeness holds vacuously.

\begin{proof}
Suppose that $W$ is \bound-addition stable, but that for some $2\leq \ell\leq k$, there exists some $c\notin W$ and $N'\subseteq N(c)$
that satisfy $q(N')\geq\ell$, but $\sum_{i\in N'}|A_i\cap W|<|N'|(\ell-1)$. The case of $\ell=1$ holds vacuously.
Observe that for every non-negative integer $x\geq0$, $(x-\ell+1)(x-\ell+2)\geq0$. Expanding and dividing this expression through by $(x+1)\ell(\ell-1)$ yields
$\frac1{x+1}\geq\frac1\ell+\frac{\ell-1-x}{\ell(\ell-1)}.$
We can use this to find that
\begin{align*}
\pav(W\cup\{c\})-\pav(W)
&\geq\sum_{i\in N'}\frac1{|A_i\cap W|+1}\\
&\geq\frac{|N'|}{\ell}
+\frac{|N'|(\ell-1)-\sum_{i\in N'}|A_i\cap W|}
{\ell(\ell-1)}\\
&\geq\frac{n}{k}+\frac1{\ell(\ell-1)}\\
&\geq\frac{n}{k}+\frac1{k(k-1)}
>\bound,
\end{align*}
where the third line uses the fact that the numerator in the second term is a positive integer. This contradicts $\bound$-addition stability, and we can thus conclude that $W$ must be $(\ell-1)$-representative.
\end{proof}

Finally, observe that the PAV score is \emph{submodular}: whenever $W\subseteq W'\subseteq C$
and $c\notin W'$, $\pav(W\cup\{c\})-\pav(W)\geq\pav(W'\cup\{c\})-\pav(W')$. Indeed, each supporter contributes $1/(|A_i\cap W|+1)$ to an addition, which decreases as the selected set grows. This allows us to conclude the following.

\begin{observation}\label{obs:PAV_addition_extension}
If $W$ is $\bound$-addition-stable for PAV, then every superset $W'\subseteq C$ is also $\bound$-addition-stable. Consequently, every completion retains the addition-side guarantees above.
\end{observation}

\subsection{Consequences of Deletion Stability}\label{sec:PAV_deletion}

The notions of \bound-addition stability, Bounded PAV improvement, and their consequences in \Cref{sec:PAV_addition} can be viewed as a guarantee against underrepresentation. Similarly, we can think of \bound-deletion stability, and its consequences, as overrepresentation axioms. These consequences include the recently introduced upper- and near-quota axioms of \citet{BLMNP26}, as well as axioms that are not ordinarily framed as preventing overrepresentation, such as affordability.

We begin by showing that deletion stability can guarantee a feasible committee by using a standard bound on the sum of marginal PAV contributions (see e.g. \citet[proofs of Theorems~6 and~10]{ABC+16a}).

\begin{lemma}\label{prop:PAV_decline_size}
    If $W$ is $\bound$-deletion-stable for PAV for $\bound>\frac{n}{k+1}$, then $|W|\leq k$.
\end{lemma}

\begin{proof}
    By $\bound$-deletion stability,
    \begin{align*}
        |W|\bound
        &\leq \sum_{c\in W}
        \bigl(\pav(W)-\pav(W\setminus\{c\})\bigr)\\
        &= \sum_{c\in W}\sum_{i\in N(c)}
        \frac{1}{|A_i\cap W|}\\
        &= \sum_{i\in N:A_i\cap W\neq\emptyset}
        \sum_{c\in A_i\cap W}\frac{1}{|A_i\cap W|}\\
        &= |\{i\in N:A_i\cap W\neq\emptyset\}|
        \leq n.
    \end{align*}
    Since $\bound>\frac{n}{k+1}$, we obtain $|W|\leq\frac{n}{\bound}<k+1$, and thus $|W|\leq k$.
\end{proof}

\noindentparagraph{Overrepresentation.}

\citet{BLMNP26} noted that their overrepresentation axioms are not satisfied by any established voting rules, aside from those introduced in their paper. Notably, this even applies to popular rules that output a partial outcome, like MES. We show that for a sufficiently high value of $\bound$, $\bound$-deletion stability implies the UQ, JUQ and JNQ overrepresentation axioms, and subsequently use this fact in \Cref{sec:rules} to formulate new voting rules that satisfy these axioms.

\begin{proposition}\label{prop:PAV_decline_upper_quota}
    $\bound$-deletion stability implies upper quota for $\bound>\frac{n}{k+1}$.
\end{proposition}

\begin{proof}
    Suppose that $W$ violates upper quota, witnessed by some $c\in W$. Then, for every $i\in N(c)$,
    $$|A_i\cap W|>
    \lceil q_c\rceil.$$
    Since $|A_i\cap W|$ is an integer, deletion stability gives
        $$\bound\leq \sum_{i\in N(c)}\frac{1}{|A_i\cap W|}
        \leq \frac{|N(c)|}{\lceil q_c\rceil+1}
        \leq \frac nk\frac{q_c}{q_c+1}
        \leq \frac{n}{k+1}
        <\bound,$$
    which is a contradiction.
\end{proof}

A slightly stronger result also holds for Justified Upper Quota.

\begin{proposition}\label{prop:PAV_deletion_JUQ}
If $W$ is $\bound$-deletion-stable
and $\bound\geq\frac{n}{k+1}$, then $W$ satisfies JUQ.
\end{proposition}

\begin{proof}
Suppose that $c\in W$, $d\notin W$, and nonempty
$N'\subseteq N(d)$ witness a JUQ violation.
Since $\bound>0$, we have $N(c)\neq\emptyset$.
Also observe that $N(c)\neq N$, else for all $i\in N$, $|A_i\cap W|>k$, and thus $|A_i\cap (W\setminus \{c\}\cup \{d\})|>k\geq \lceil q(N')\rceil$ for all voters $i\in N'$.
The calculation in the proof of
\Cref{prop:PAV_decline_upper_quota} therefore gives
$$
\bound
\leq\frac nk\frac{q_c}{q_c+1}
<\frac{n}{k+1},
$$
a contradiction. Here, the last inequality is strict because $|N(c)|<n$.
\end{proof}

For near quota as well as our upcoming priceability result, we require the following auxiliary argument that allows for an inverse-polynomial gap between our $\bound$ bound and $\nicefrac{n}{k}$, which we will later use to establish polynomial computability.

\begin{lemma}\label{lem:PAV_deletion_support}
If $W$ is $\bound$-deletion-stable
for $\bound\geq \frac{n}{k}-\frac{1}{k(k+1)}$, and $|W|\leq k$ then for every nonempty $C'\subseteq W$ we must have
$q(\{i\in N:A_i\cap C'\neq\emptyset\})\geq |C'|$.
Moreover, every $c\in W$ has a supporter $i\in N(c)$ with
$|A_i\cap W|\leq q_c$.
\end{lemma}

\begin{proof}

First, observe that for integers $a,b$ with $1\leq b\leq k$,
$$
\frac ab<\frac nk
\quad\Longrightarrow\quad
\frac ab\leq\frac nk-\frac1{kb}
\leq\frac nk-\frac1{k^2}<\bound.
$$

A similar argument is used in \citet[Lemma 1]{CaEl26a} to obtain a local search algorithm for PAV that satisfies EJR+ for the Droop quota.

For the first conclusion, we find that for nonempty $C'\subseteq W$, $\bound$-deletion stability gives
\begin{align*}
\bound
&\leq\frac1{|C'|}\sum_{c\in C'}
  \bigl(\pav(W)-\pav(W\setminus\{c\})\bigr)\\
&=\frac1{|C'|}\sum_{i:A_i\cap C'\neq\emptyset}
  \frac{|A_i\cap C'|}{|A_i\cap W|}\\
&\leq\frac{|\{i\in N:A_i\cap C'\neq\emptyset\}|}{|C'|}.
\end{align*}

Thus, since $\frac{|\{i\in N:A_i\cap C'\neq\emptyset\}|}{|C'|}\geq \bound$, it must also be the case that $|\{i\in N:A_i\cap C'\neq\emptyset\}|\geq \frac{n}{k}|C'|$.

For the second conclusion, we know that by $\bound$-deletion stability, for any $c\in W$, we have $N(c)\neq\emptyset$ and $\sum_{i\in N(c)}\frac1{|A_i\cap W|}\geq \bound$. Thus, for some $i\in N(c)$, $\frac{|N(c)|}{|A_i\cap W|}\geq \bound$, and thus $\frac{|N(c)|}{|A_i\cap W|}\geq \frac{n}{k}$, by the inequality argument above. Rearranging gives $|A_i\cap W|\leq q_c$.
\end{proof}

\begin{corollary}\label{prop:PAV_deletion_JNQ}
If $W$ is $\bound$-deletion-stable for $\bound\geq \frac{n}{k}-\frac{1}{k(k+1)}$, then $W$ satisfies JNQ.
\end{corollary}

\begin{proof}
Observe that, if $n=1$, then JNQ holds vacuously, as the sole voter's satisfaction cannot be simultaneously strictly above and strictly below $k$. If $n>1$, then $\bound\geq \frac{n}{k}-\frac{1}{k(k+1)}>\frac{n}{k+1}$ and thus by \Cref{prop:PAV_decline_size} $|W|\leq k$. In this case we can directly apply \Cref{lem:PAV_deletion_support} to rule out the overrepresentation condition of a JNQ violation, since every $c\in W$ has a supporter $i\in N(c)$ with
$|A_i\cap W|\leq q_c$, and thus $|A_i\cap W|-q_c\leq \frac12\indicator[d\notin A_i]$.
\end{proof}

\noindentparagraph{Affordability.} We now show that $\bound$-deletion stability can guarantee affordability, and briefly consider the consequences of this result.

\begin{corollary}\label{prop:PAV_deletion_affordability}
If $W$ is $\bound$-deletion-stable for $\bound> \frac{n}{k}-\frac{1}{k(k+1)}$, then $W$ is affordable.
\end{corollary}

\begin{proof}
By \bound-deletion-stability, \Cref{prop:PAV_decline_size} gives $|W|\leq k$, which means that the first conclusion of \Cref{lem:PAV_deletion_support} holds. We can then apply the fractional form of Hall's theorem to show that $W$ must satisfy affordability; see \citet[proof of Lemma~2]{PPS21b} for the payment construction, or \citet[Theorem~7]{CeSt21a} for the equivalent maximin-support characterization.
\end{proof}

For $\bound\geq n/k$, we can give the affordable payment system explicitly, without invoking Hall's theorem.

\begin{proposition}\label{prop:PAV_simple_payments}
Suppose that $W$ is $\bound$-deletion-stable for PAV with $\bound\geq \frac{n}{k}$. Set $$p_i(c)=\frac{1}{|A_i\cap W|(\pav(W)-\pav(W\setminus\{c\}))}$$ for all $i\in N$ and all $c\in A_i\cap W$, and $p_i(c)=0$ otherwise. The payment system $(p_i)_{i\in N}$ is affordable.
\end{proposition}

The payment system $(p_i(c))_{i\in N}$ can be thought of as each voter $i\in N$ paying for each approved candidate $c\in A_i\cap W$ their marginal PAV contribution, $\nicefrac{1}{|A_i\cap W|}$, normalized by the overall marginal PAV score of that candidate. 

EP-PAV meets the deletion stability threshold of $\nicefrac{n}{k}$ required by the formula above. With the lower threshold used by LP-PAV, the committee is still affordable, but the payments of \Cref{prop:PAV_simple_payments} can exceed some voter's budget, as shown in \Cref{prop:LD_PAV_simple_payments_fail}.

\begin{proof}
Observe that for $c\in W$, deletion stability gives
$$\pav(W)-\pav(W\setminus\{c\})
\geq\bound\geq\frac nk>0,$$
and thus our payments are well defined. Conditions 1 and 4 of affordability hold immediately.

For condition 3, note that $\pav(W)-\pav(W\setminus\{c\})=\sum_{i\in N(c)}\frac1{|A_i\cap W|}$, and thus $\sum_{i\in N}p_i(c)=1$ for all $c\in W$.

For condition 2, each represented voter $i\in N$ with $A_i\cap W\neq \emptyset$ pays
$$
\sum_{c\in C}p_i(c)
\leq
\sum_{c\in A_i\cap W}\frac{k}{n|A_i\cap W|}
=\frac kn,
$$
and unrepresented voters pay nothing.
\end{proof}

Affordability also gives coverage guarantees by the completion arguments for affordable committees detailed by \citet[Corollary~12 and Theorem~14]{BrPe24a}. Corresponding utilitarian guarantees are discussed in \Cref{sec:utilitarian_guarantees}.

\begin{corollary}\label{cor:PAV_CC_completion}
If $W$ is $\bound$-deletion-stable for $\bound>\frac{n}{k}-\frac{1}{k(k+1)}$
then its CC completion has representation ratio at least
$3/4$, and its sequential CC completion has representation
ratio at least $1-1/e$.
\end{corollary}

Starting from an affordable committee $W$, we can run sequential Phragm\'en with each voter's initial balance set to minus their total payment for $W$, using only positive balances to fund new candidates \citep{PPS21b,PeSk20b}. Once each voter has accrued $\nicefrac{k}{n}$ units of credit, the resulting committee is a priceable extension at budget $B=k$. Continuing until all $k$ seats are filled gives a priceable completion at a possibly larger budget.

\begin{corollary}\label{cor:PAV_priceable_completion}
If $W$ is $\bound$-deletion-stable for $\bound>\frac{n}{k}-\frac{1}{k(k+1)}$, then it admits a priceable superset at budget $B=k$ and a priceable completion at a possibly larger budget.
\end{corollary}

\noindentparagraph{Laminar Proportionality.}

On the laminar domain, the following result combines our Upper Quota guarantee with the laminar proportionality of MES established by \citet[Theorem~1]{PeSk20b}.

\begin{proposition}\label{prop:PAV_laminar_completion}
On a laminar instance, let $W$ be $\bound$-deletion-stable with
$\bound>n/(k+1)$.
If there are no $c\in W$ and $d\notin W$ such that
$
N(c)\subsetneq N(d),
$
then $W$ has a laminar-proportional completion.
Such a completion is obtained by running MES, starting from the empty committee, tiebreaking in favor of candidates in $W$.
\end{proposition}

\begin{proof}
Observe that by \Cref{prop:PAV_decline_upper_quota}, $W$ satisfies Upper Quota.

Run MES with ties favoring members of $W$, and let $W^*$
be its outcome. By \citet[Theorem~1 and its proof]{PeSk20b},
$W^*$ is laminar proportional, every voter spends $k/n$,
and each selected candidate's cost is shared equally
among its supporters.

Suppose that there exists some $c\in W\setminus W^*$. Laminar proportionality implies that every candidate $c'\in C$ with $N(c')\supset N(c)$ must have been selected in $W^*$. By MES, with our stated tiebreaking, there cannot be a candidate $c'\in W^*\setminus W$ with $N(c')\subseteq N(c)$. Thus, it must be the case that for every $i\in N(c)$, $A_i\cap W^*\subseteq (A_i\cap W)\setminus\{c\}$.

Thus, for each such voter,
$$
\frac{k}{n}
=\sum_{c'\in A_i\cap W^*}\frac1{|N(c')|}
\leq\frac{|A_i\cap W^*|}{|N(c)|}.
$$
Hence
$$
|A_i\cap W|
\geq |A_i\cap W^*|+1
\geq\lceil q_c\rceil+1
\qquad\text{for every }i\in N(c),
$$
contradicting UQ. Thus $W\subseteq W^*$.
\end{proof}

\subsection{Joint Guarantees}\label{sec:PAV_joint}

We now investigate properties requiring some level of addition and deletion stability simultaneously.

\noindentparagraph{Utilitarian Guarantees.}\label{sec:utilitarian_guarantees}

In their work, \citet{BrPe24a}.\footnote{With a correction in \citet{BBP26b}} show that MES, completed with AV, achieves a utilitarian guarantee of $\nicefrac{2}{\sqrt{k}}-\nicefrac{2}{k}.$ We show that this also holds for any committee satisfying a particular level of addition and deletion stability.

\begin{proposition}\label{prop:PAV_utilitarian_completion}
If $W$ is $\bound_1$-addition-stable with $\bound_1\leq\frac nk$, and $\bound_2$-deletion-stable with $\bound_2>\frac nk-\frac1{k(k+1)}$, then its AV completion has a utilitarian ratio of at least $\nicefrac{2}{\sqrt{k}}-\nicefrac{2}{k}$.
\end{proposition}

Unlike \Cref{cor:PAV_CC_completion}, this does not follow directly from the results of \citet{BrPe24a}. Deletion stability gives a lower bound on the welfare contributed by
each candidate selected in $W$. Meanwhile, addition stability tells us something about omitted candidates; it gives a lower bound on how many members of $W$ the supporters of the most popular candidate omitted by the AV completion approve on average. This gives a further contribution to the welfare bound. The full argument is in \Cref{app:PAV_nearquota_welfare_proof}. Curiously, the same comparison gives a utilitarian ratio $\nicefrac{2}{\sqrt{k}}-\nicefrac{2}{k}$ for standard PAV and a polynomial-time local search variant; see \Cref{cor:regular_PAV_utilitarian} in the appendix.

\noindentparagraph{Exact Lower Quota.}
On apportionment instances, PAV coincides with D'Hondt's method, which satisfies lower quota \citep{BLS18a}. The following stability conditions ensure that every party receives exactly its lower quota.

\begin{corollary}\label{prop:exact_lower_quota}
On an apportionment instance, if $W$ is $\bound_1$-addition-stable with $\bound_1<\frac nk$ and $\bound_2$-deletion-stable with $\bound_2>\frac nk-\frac1{k(k+1)}$, then $W$ satisfies exact lower quota.
\end{corollary}

\begin{proof}
By \Cref{prop:PAV_improvement_EJR} and \Cref{prop:PAV_deletion_affordability}, $W$ satisfies EJR+ and is affordable. Since each party has $k$ candidates, EJR+ gives $|W\cap C_j|\geq\lfloor q(N_j)\rfloor$ for every $j$. Only voters in $N_j$ can fund candidates in $C_j$, and their total budget is $q(N_j)$. Thus affordability gives $|W\cap C_j|\leq q(N_j)$, and integrality yields equality with $\lfloor q(N_j)\rfloor$.
\end{proof}

\noindentparagraph{Local Search.} We show that an outcome that is addition and deletion stable can be found in polynomial time for an appropriate choice of bounds.

\begin{proposition}\label{prop:PAV_local_search}
For rational $\bound_1>\bound_2\geq0$, a committee
that is $\bound_1$-addition-stable and
$\bound_2$-deletion-stable, and has no
$c\in W$, $d\notin W$ with $N(c)\subsetneq N(d)$,
can be found in time polynomial in the input size and
$1/(\bound_1-\bound_2)$.
\end{proposition}

\begin{proof}
We adapt the bounded improvement local-search argument of \citet[Proposition~1]{AEH+18a} to the penalized PAV objective.
Set
$\bound=\nicefrac{\bound_1+\bound_2}{2}.$
Following \Cref{def:LD_PAV}, starting from $W=\emptyset$, repeatedly perform any applicable
move:
\begin{enumerate}
\item Add $c\notin W$ if
$\pav(W\cup\{c\})-\pav(W)>\bound_1$.
\item Delete $c\in W$ if
$\pav(W)-\pav(W\setminus\{c\})<\bound_2$.
\item Replace $c\in W$ by $d\notin W$ if
$N(c)\subsetneq N(d)$.
\end{enumerate}
Note that the final step is only necessary for the guarantee in \Cref{prop:PAV_laminar_completion}.

Consider the potential $\apav(W)=\pav(W)-\bound|W|$.
Each addition or deletion increases it by more than
$\nicefrac{\bound_1-\bound_2}{2}$.
Each swap increases
it by at least $\nicefrac{1}{|C|}$.
The potential starts at zero and always satisfies
$$
\apav(W)\leq\pav(W)
\leq n\sum_{j=1}^{|C|}\frac1j
\leq n(1+\log|C|).
$$
Consequently, there are at most $2n(1+\log|C|)/(\bound_1-\bound_2)$ additions and deletions, and at most $|C|n(1+\log|C|)$ swaps. Each iteration checks at most $|C|$ additions and deletions and $|C|^2$ support-containment pairs. The marginal scores and potential comparisons use rational numbers of polynomial bit length, proving the claim.
\end{proof}

\subsection{Combining Our Guarantees}\label{sec:rules}

Collating the results derived in previous sections, we show that our two penalized PAV rules satisfy a selection of attractive properties.

\begin{corollary}\label{cor:PAV_polynomial_guarantees}
LP-PAV (\Cref{def:LD_PAV}) with
$$
\bound_1=\frac nk-\frac1{3k(k+1)},
\qquad
\bound_2=\frac nk-\frac2{3k(k+1)}
$$
runs in polynomial time and returns a feasible committee $W$ with the following properties:
\begin{itemize}
\item $W$ is affordable, and can be extended to a priceable committee. \item $W$ satisfies EJR+, $2$-frugal Lindahl priceability, $(\ell-1)$-representativeness, UQ, JUQ, and JNQ.
\item Its AV completion has a utilitarian guarantee of $\nicefrac{2}{\sqrt{k}}-\nicefrac{2}{k}$. Its CC completion has a representation guarantee of $\nicefrac{3}{4}$.
\item On apportionment instances, $W$ satisfies exact lower quota. On laminar instances, $W$ has a completion that is laminar proportional.
\end{itemize}
\end{corollary}

We also consider exact local maximization using the additions, deletions, and support-dominance swaps of
\Cref{prop:PAV_local_search}, using $\bound=\bound_1=\bound_2$. For additions, we implement maximum-cardinality tiebreaking, by adding $c\notin W$ if $\pav(W\cup\{c\})-\pav(W)\geq \bound_1$. Global maximization of the $\apav$ objective, with maximum cardinality tiebreaking, implies all three conditions.

\begin{observation}\label{obs:PAV_support_dominance}
Let $\bound>0$. Every exact global or local maximizer $W$ of $\apav$, with maximum cardinality tiebreaking, is $\bound$-deletion-stable. $W$ satisfies $\pav(W\cup\{c\})-\pav(W)<\bound$ for every $c\notin W$, and thus $W$ is $\bound_1$-addition-stable for some $\bound_1<\bound$. Moreover, there are no $c\in W$, $d\notin W$ with $N(c)\subsetneq N(d)$.
\end{observation}

In particular maximum-cardinality tiebreaking allows our results for EJR+ (\Cref{prop:PAV_improvement_EJR}) and exact lower quota (\Cref{prop:exact_lower_quota}) to include the $\bound=\nicefrac{n}{k}$ endpoint, as shown in \Cref{fig:PAV_stability_ranges}.

\begin{corollary}\label{cor:PAV_exact_rule}
EP-PAV with $\bound=\nicefrac{n}{k}$ satisfies all guarantees in \Cref{cor:PAV_polynomial_guarantees}, except polynomial-time computability. Additionally its outcome admits the explicit affordable payment system of \Cref{prop:PAV_simple_payments}.
\end{corollary}

Finally, we observe a curious connection between the standard PAV rule, and EP-PAV; standard PAV gives a size-$k$ committee that is stable at a common addition and deletion threshold.

\begin{proposition}\label{prop:PAV_size_k_local_optimum}
Every instance admits some $0\leq\bound\leq \nicefrac{n}{k+1}$ and a size-$k$ committee that is $\bound$-stable.
\end{proposition}

\begin{proof}
Let $W$ be the output of regular PAV. Thus, for any $c\in W$, $d\notin W$ we have $\pav(W)\geq \pav(W\cup \{d\}\setminus \{c\})$. By submodularity of $\pav$ we have $\pav(W\cup\{d\}\setminus \{c\})-\pav(W\setminus \{c\})\geq\pav(W\cup\{d\})-\pav(W)$. Combining these, we can obtain $\pav(W)-\pav(W\setminus \{c\})\geq \pav(W\cup \{d\})-\pav(W)$. 

If $W=C$, take $\bound=0$. Otherwise, let $\bound=\max\{\pav(W\cup \{d\})-\pav(W):d\notin W\}$ be its largest addition gain, or zero if $C=W$. Since $W$ is a PAV winner, by \citet[proof of Proposition~D.2]{GoYa26a}, $\bound\leq \nicefrac{n}{k+1}$.

Combining these observations, for any $c\in W$, $d\notin W$ we have $\pav(W)-\pav(W\setminus \{c\})\geq \bound \geq \pav(W\cup \{d\})-\pav(W)$ and thus $W$ is $\bound$-stable.
\end{proof}

\section{Conclusion}

\begin{table*}[t]
\defcitealias{AEH+18a}{AEH+18}
\defcitealias{PeSk20b}{PS20}
\defcitealias{BrPe23a}{BP23}
\defcitealias{Skow26b}{S26b}
\defcitealias{GoYa26a}{GY26}
\defcitealias{BLMNP26}{BLM+26}
\defcitealias{BrPe24a}{BP24}
\defcitealias{BBP26b}{BBP26}
\defcitealias{LaSk20b}{LS20}
\defcitealias{LaSk22a}{LS23}

\newcommand{\tabproved}{\cellcolor{green!12}}
\newcommand{\tabresult}[1]{\tabproved\hyperref[#1]{[\Cref*{#1}]}}

\centering
\footnotesize
\setlength{\tabcolsep}{4pt}
\renewcommand{\arraystretch}{1.2}
\setlength{\aboverulesep}{0pt}
\setlength{\belowrulesep}{0pt}

\begin{tabularx}{\textwidth}{@{}>{\raggedright\arraybackslash}X|lll@{}}
\toprule
 & PAV & Penalized PAV & MES \\
\midrule

Poly-time computation
  & Local search \citepalias{AEH+18a}
  & Local search \tabresult{prop:PAV_local_search}
  & Yes \citepalias{PeSk20b} \\

\midrule

EJR+
  & Yes \citepalias{BrPe23a}
  & Yes \tabresult{prop:PAV_improvement_EJR}
  & Yes \citepalias{BrPe23a} \\

Strong EJR+
  & No \tabresult{prop:comparison_counterexamples}
  & No \tabresult{prop:comparison_counterexamples}
  & Yes \citepalias{Skow26b} \\

Representativeness
  & $\ell-1$ \citepalias{BrPe23a}
  & $\ell-1$ \tabresult{prop:PAV_improvement_representativeness}
  & $\nicefrac{(\ell-1)}{2}$ \citepalias{BrPe23a} \\

$2$-frugal Lindahl priceability
  & Yes \citepalias{GoYa26a}
  & Yes \tabresult{prop:PAV_improvement_frugal_Lindahl}
  & No \citepalias{PeSk20b} \\

\midrule

Upper quota
  & No \citepalias{BLMNP26}
  & Yes \tabresult{prop:PAV_decline_upper_quota}
  & PER elimination \citepalias{BLMNP26} \\

Justified near quota
  & No \citepalias{BLMNP26}
  & Yes \tabresult{prop:PAV_deletion_JNQ}
  & PER elimination \citepalias{BLMNP26} \\

Affordability
  & No \citepalias{PeSk20b}
  & Yes \tabresult{prop:PAV_deletion_affordability}
  & Yes \citepalias{PeSk20b} \\

Priceability
  & No \citepalias{PeSk20b}
  & Phragm\'en completion \tabresult{cor:PAV_priceable_completion}
  & Yes \citepalias{PeSk20b} \\

Laminar proportionality
  & No \citepalias{PeSk20b}
  & MES completion \tabresult{prop:PAV_laminar_completion}
  & Yes \citepalias{PeSk20b} \\

\midrule

Utilitarian ratio
  & $\frac{2}{\sqrt{k}}-\frac{2}{k}$
    \tabresult{cor:regular_PAV_utilitarian}
  & $\frac{2}{\sqrt{k}}-\frac{2}{k}$
    \tabresult{prop:PAV_utilitarian_completion}
  & $\frac{2}{\sqrt{k}}-\frac{2}{k}$
    \citepalias{BrPe24a,BBP26b} \\
(with AV completion)
  & \tabproved & \tabproved & \\

Representation ratio
  & $\nicefrac{1}{2}$ \citepalias{LaSk20b}
  & $\nicefrac{3}{4}$ \tabresult{cor:PAV_CC_completion}
  & $\nicefrac{3}{4}$ \citepalias{BrPe24a} \\
(with CC completion)
  & & \tabproved & \\

\midrule

Pareto-optimality
  & Yes (Exact) \citepalias{LaSk22a}
  & No \tabresult{prop:comparison_counterexamples}
  & No \citepalias{LaSk22a} \\

\bottomrule
\end{tabularx}

\caption{Comparison of properties satisfied by PAV, P-PAV, and MES.
Shaded cells refer to results proved in this paper.}
\label{tab:comparison}
\end{table*}

Our penalized PAV rules combine score optimization with the flexibility of a partial outcome. The analysis separates the roles of addition and deletion stability, treating them as parameterized axioms of underrepresentation and overrepresentation, respectively. When these stability notions are combined, thresholds close to $\nicefrac{n}{k}$ give affordability and strong proportionality guarantees. Both EP-PAV and LP-PAV achieve these guarantees, with LP-PAV running in polynomial time. Different completions provide welfare and coverage guarantees, as well as laminar proportionality on laminar instances.

Our rules produce outcomes that are axiomatically much closer to MES than to standard PAV, as can be seen in \Cref{tab:comparison}, which summarizes the main properties of all three rules.  The primary difference is with respect to lower quota proportionality: MES satisfies Strong EJR+, while penalized PAV guarantees $(\ell-1)$-representativeness. Several other differences can be addressed through suitable completion or elimination procedures.

Deepening our understanding of simpler multiwinner voting rules, like PAV, can serve as a blueprint for investigating more complex ones. This suggests using our analysis of penalized PAV as inspiration for examining the recently introduced harmonic entropy rule \citep{BGP26a}. Considering the consequences of deletion stability for harmonic entropy could lead to further insights about overrepresentation axioms and favorable completions in that setting. Another possible direction for future work would be an empirical comparison of MES, penalized PAV, and related rules. We have established that these are closely connected in terms of their theoretical guarantees. However, such worst-case guarantees alone give an incomplete picture of empirical performance in committee elections and participatory budgeting \citep{LaSk20b,GPS+24a}. Thus, it would be of interest to compare penalized PAV to established rules with respect to proportionality, welfare, coverage, and other properties provided by their outcomes, and appropriate completions.

\section*{Acknowledgments and AI Use}\label{sec:acks}

I thank Markus Brill and Matthias Greger for insightful discussions.

During an investigation for another project, ChatGPT Sol 6.1 proposed a penalized modification of the harmonic entropy rule and an argument that it yields affordable outcomes satisfying Core+, adapting results of \citet{BGP26a}. I obtained this derivation on October 16, 2026, independently of the related result of \citet{AzVo26a}, which became publicly available on October 28, 2026. This observation motivated the investigation of an analogous modification to PAV, and the related results presented in this paper.

The research directions, questions, and conceptual framework of this paper are my own. These include separating the roles of addition and deletion stability, their visualization in \Cref{fig:PAV_stability_ranges}, and the comparison of my penalized PAV rules to the Method of Equal Shares. I also used ChatGPT to assist with literature searches and developing and checking candidate proofs and counterexamples. Parts of the paper were initially drafted by ChatGPT, but these have been extensively edited and improved, and I take full responsibility for its contents.

\bibliographystyle{ACM-Reference-Format}
\bibliography{abb,algo,refs}

\newpage
\appendix

\appendix

\section{Utilitarian guarantees}\label{app:PAV_utilitarian_guarantees}
\label{app:preliminaries}

Let $\overline W$ be an AV completion of $W$, and let $W^*$ be a size-$k$ committee maximizing utilitarian welfare. Following the welfare comparison of \citet[Section~4.1]{BrPe24a}, we first bound the welfare lost when an AV completion omits a popular candidate.

\subsection{A shared welfare comparison}

\begin{lemma}\label{lem:PAV_utilitarian_comparison}
Let $\bound_1\geq\bound_2\geq0$. If $|W|\leq k$ and $W$ is
$\bound_1$-addition-stable and $\bound_2$-deletion-stable for PAV, then
\[
    \operatorname{sw}(\overline W)
    \geq \left(\frac2{\sqrt{k}}-\frac2k\right)\operatorname{sw}(W^*)-k(\bound_1-\bound_2).
\]
\end{lemma}

\begin{proof}
If $\overline W=C$, the claim is immediate. Otherwise, let $c^*$ have the most supporters among the candidates outside $\overline W$. Every candidate added by AV has at least $|N(c^*)|$ supporters. By $\bound_2$-deletion stability, each candidate in $W$ has at least $\bound_2$ supporters, since each supporter contributes at most $1$ to its marginal PAV score. 

If $|N(c^*)|\leq\bound_1$, then replacing a selected candidate by an omitted one increases $\operatorname{sw}$ by at most $\bound_1-\bound_2$. We can reach $W^*$ from $\overline W$ by performing at most $k$ such swaps, which means that $\operatorname{sw}(\overline W)\geq \operatorname{sw}(W^*)-k(\bound_1-\bound_2)$, proving our claim.

Thus, we suppose that $|N(c^*)|>\bound_1>0$. Put $\alpha=\nicefrac{|N(c^*)|}{\bound_1}$ and let $C'$ consist of all candidates in $C$ with more supporters than $c^*$. From our choice of $c^*$ we know that $C'\subseteq \overline W$. Because every candidate in $C\setminus C'$ has at most $|N(c^*)|$
supporters,
$$
   \operatorname{sw}(W^*)\leq\operatorname{sw}(C')+(k-|C'|)|N(c^*)|.
$$
We consider two cases.

\emph{Case 1:} Suppose $|C'|\geq\alpha-1$.
Recall that by $\bound_2$-deletion stability, any $c\in W\setminus C'$ must have $|N(c)|\geq \bound_2$. This also holds for any $c\in\overline W\setminus W$, since each
candidate added by AV has at least $|N(c^*)|>\bound_1\geq\bound_2$
supporters. Using this,
\begin{align*}
 \operatorname{sw}(\overline W\setminus C')+|C'||N(c^*)|
 &\geq (k-|C'|)\bound_2+|C'||N(c^*)|\\
 &=k\lambda_1 +|C'||N(c^*)| -|C'|\bound_2 -k\lambda_1 +k\bound_2 \\
 &\geq k\lambda_1 +\alpha\bound_1|C'| -|C'|\bound_1 -k\lambda_1 +k\bound_2 \\
 &=\bound_1\bigl(k+|C'|(\alpha-1)\bigr)-k(\bound_1-\bound_2)\\
 &\geq \bound_1(k+\alpha^2-2\alpha)-k(\bound_1-\bound_2).
\end{align*}

\emph{Case 2:} Suppose $|C'|<\alpha-1$.
For each $c\in W\setminus C'$, split its welfare contribution into
its marginal PAV score and the remainder:
\begin{align*}
|N(c)|
&= \sum_{i\in N(c)}\frac{1}{|A_i\cap W|}
   + \sum_{i\in N(c)}\left(1-\frac{1}{|A_i\cap W|}\right)\\
&\geq \bound_2
   + \sum_{i\in N(c)\cap N(c^*)}
      \left(1-\frac{1}{|A_i\cap W|}\right).
\end{align*}
The inequality uses our $\bound_2$-deletion stability assumption. Each candidate added by AV has at least $|N(c^*)|>\bound_1\geq\bound_2$ supporters. Summing over $\overline W\setminus C'$ gives
$$
\operatorname{sw}(\overline W\setminus C')
\geq (k-|C'|)\bound_2 +\sum_{\substack{i\in N(c^*)\\ A_i\cap W\neq\emptyset}}
 |A_i\cap(W\setminus C')|
 \left(1-\frac{1}{|A_i\cap W|}\right).
$$

To bound this sum, we first show that all voters in $N(c^*)$ approve at least $\alpha-1$ members of $W$, on average. Our addition stability assumption bounds the marginal PAV gain of adding $c^*$ to $W$ by $\bound_1$, so the arithmetic-harmonic mean inequality gives
$$
 \frac{1}{|N(c^*)|}\sum_{i\in N(c^*)} (|A_i\cap W|+1)
 \geq
 \frac{|N(c^*)|}{\displaystyle\sum_{i\in N(c^*)}1/(|A_i\cap W|+1)}
 \geq \frac{|N(c^*)|}{\bound_1}=\alpha.
$$

We now bound the remaining welfare from the supporters of $c^*$. Let $f(x)=(x-|C'|)(1-1/x)$. Since at most $|C'|$ of the winners a voter approves lie in $C'$, for each $i\in N(c^*)$ with $A_i\cap W\neq\emptyset$ we have
$$
 |A_i\cap(W\setminus C')|
 \left(1-\frac{1}{|A_i\cap W|}\right)
 \geq f(|A_i\cap W|).
$$
Voters who approve no winner contribute zero, which equals $f(1)$.

The derivative $f'(x)=1-|C'|/x^2$ is nondecreasing for $x>0$,
so $f$ is convex. The derivative is also nonnegative for $x\geq\alpha-1$,
since either $C'=\emptyset$ or $\alpha-1>|C'|\geq1$.
Thus, applying Jensen's inequality gives
\begin{align*}
 &\frac{1}{|N(c^*)|}
 \sum_{\substack{i\in N(c^*)\\ A_i\cap W\neq\emptyset}}
 |A_i\cap(W\setminus C')|
 \left(1-\frac{1}{|A_i\cap W|}\right)\\
 &\quad\geq \frac{1}{|N(c^*)|}
 \sum_{i\in N(c^*)} f\bigl(\max\{1,|A_i\cap W|\}\bigr)\\
 &\quad\geq
 f\left(\frac{1}{|N(c^*)|}
 \sum_{i\in N(c^*)}\max\{1,|A_i\cap W|\}\right)
 \geq f(\alpha-1). \tag{$*$}\label{eq:PAV_Jensen_welfare}
\end{align*}
The last inequality uses our earlier bound on the average number
of approved winners and the fact that $f$ is nondecreasing on
$[\alpha-1,\infty)$. Combining this with our earlier welfare bound gives
$$
 \operatorname{sw}(\overline W\setminus C')
\geq (k-|C'|)\bound_2+|N(c^*)|(\alpha-1-|C'|)\left(1-\frac{1}{\alpha-1}\right).
$$
Using $|N(c^*)|=\alpha\bound_1$ gives
\begin{align*}
 \operatorname{sw}(\overline W\setminus C')+|C'|\,|N(c^*)|
 &\geq (k-|C'|)\bound_2+|N(c^*)|\left[|C'|+(\alpha-1-|C'|)\left(1-\frac{1}{\alpha-1}\right)\right]\\
 &= (k-|C'|)\bound_2
  +|N(c^*)|\left(\alpha-2+\frac{|C'|}{\alpha-1}\right)\\
 &= \bound_1\left(k+\alpha^2-2\alpha+\frac{|C'|}{\alpha-1}\right)
  -(k-|C'|)(\bound_1-\bound_2)\tag{$**$}\label{eq:PAV_refined_welfare_comparison}\\
   &\geq \bound_1\left(k+\alpha^2-2\alpha\right)
  -k(\bound_1-\bound_2).
\end{align*}

Observe that both cases give the same bound.
Then, using the AM-GM inequality for $k+\alpha^2\geq 2\alpha\sqrt{k}$ we obtain:
\begin{align*}
 \operatorname{sw}(\overline W)-\operatorname{sw}(C')
     +|C'|\,|N(c^*)|
 &\geq \bound_1(k+\alpha^2-2\alpha)-k(\bound_1-\bound_2)\\
 &\geq \left(\frac{2}{\sqrt{k}}-\frac{2}{k}\right)
     k|N(c^*)|-k(\bound_1-\bound_2).
     \tag{$***$}\label{eq:PAV_adjusted_welfare}
\end{align*}
Recall that
$$
 \operatorname{sw}(W^*)
 \leq \operatorname{sw}(C')+(k-|C'|)|N(c^*)|.
$$
The right-hand side is $k|N(c^*)|$ plus the nonnegative quantity
$\operatorname{sw}(C')-|C'|\,|N(c^*)|$. Since
$0\leq 2/\sqrt{k}-2/k\leq 1$, restoring this quantity to the
preceding lower bound gives
\begin{align*}
 \operatorname{sw}(\overline W)
 &\geq \left(\frac{2}{\sqrt{k}}-\frac{2}{k}\right)
 \bigl(\operatorname{sw}(C')+(k-|C'|)|N(c^*)|\bigr)
 -k(\bound_1-\bound_2)\\
 &\geq \left(\frac{2}{\sqrt{k}}-\frac{2}{k}\right)
 \operatorname{sw}(W^*)-k(\bound_1-\bound_2).
\end{align*}

\end{proof}

\begin{corollary}\label{lem:PAV_common_threshold_welfare}
If $|W|\leq k$ and $W$ is $\bound$-stable for PAV, its AV completion
has utilitarian ratio at least $2/\sqrt{k}-2/k$.
\end{corollary}
\begin{proof}
Set $\bound_1=\bound_2=\bound$ in \Cref{lem:PAV_utilitarian_comparison}.
\end{proof}

\subsection{Penalized PAV}\label{app:PAV_nearquota_welfare_proof}

\begin{proof}[Proof of \Cref{prop:PAV_utilitarian_completion}]
Write $r=\nicefrac{2}{\sqrt{k}}-\nicefrac{2}{k}$ for the desired
ratio. By \Cref{prop:PAV_deletion_affordability}, $W$ is affordable.
Since each voter has budget $\nicefrac{k}{n}$, each member of $W$
has at least $\nicefrac{n}{k}$ supporters, and $W$ represents at
least $\nicefrac{n|W|}{k}$ voters \citep[Observation~1]{BrPe24a}.

If $\overline W=C$, the claim is immediate. Otherwise, choose $c^*$
as in \Cref{lem:PAV_utilitarian_comparison}. If
$|N(c^*)|\leq\nicefrac{n}{k}$, every member of $\overline W$ has
at least as many supporters as every omitted candidate, so
$\overline W$ maximizes welfare. Thus, suppose $|N(c^*)|>\nicefrac{n}{k}$. Since $W$ is also $\nicefrac{n}{k}$-addition-stable, we apply the shared comparison
with $\bound_1=\nicefrac{n}{k}$. With this choice, $\alpha=\nicefrac{|N(c^*)|}{\bound_1}=q_{c^*}>1$. As before, let $C'\subseteq W$ consist of the candidates with more supporters than $c^*$.

\emph{Case 1:} Suppose $C'=\emptyset$. We split $\operatorname{sw}(W)$
by counting the first approved winner of each represented voter
separately from their remaining approved winners:
\[
 \operatorname{sw}(W)
 = |\{i\in N:A_i\cap W\neq\emptyset\}|
   + \sum_{i\in N}\max\{0,|A_i\cap W|-1\}.
\]
By affordability, the first term is at least $\nicefrac{n|W|}{k}$.
Since $C'=\emptyset$, each represented voter contributes
$|A_i\cap W|-1$ to the sum in \eqref{eq:PAV_Jensen_welfare},
and $f(\alpha-1)=\alpha-2$. Multiplying
\eqref{eq:PAV_Jensen_welfare} by $|N(c^*)|$ therefore gives
\[
 \sum_{i\in N}\max\{0,|A_i\cap W|-1\}
 \geq \sum_{i\in N(c^*)}\max\{0,|A_i\cap W|-1\}
 \geq |N(c^*)|(\alpha-2).
\]

Each candidate added by AV has at least $|N(c^*)|>\nicefrac{n}{k}$
supporters. Thus
\begin{align*}
 \operatorname{sw}(\overline W)
 &= \operatorname{sw}(W)+\operatorname{sw}(\overline W\setminus W)\\
 &\geq \frac{n|W|}{k}+|N(c^*)|(\alpha-2)
       +(k-|W|)\frac{n}{k}\\
 &= \frac{n}{k}(k+\alpha^2-2\alpha)
 \geq r k|N(c^*)|,
\end{align*}
where the last step uses AM--GM as in the shared comparison.
Since $C'=\emptyset$, we have $\operatorname{sw}(W^*)\leq k|N(c^*)|$,
which proves the claim.

\emph{Case 2:} Suppose $C'\neq\emptyset$. Apply
\eqref{eq:PAV_adjusted_welfare} with $\bound_1=\nicefrac{n}{k}$
and $\bound_2=\nicefrac{n}{k}-\nicefrac{1}{k(k+1)}$. Adding back
$\operatorname{sw}(C')-|C'|\,|N(c^*)|$ gives
\[
 \operatorname{sw}(\overline W)
 \geq rk|N(c^*)|
 +\bigl(\operatorname{sw}(C')-|C'|\,|N(c^*)|\bigr)
 -\frac{1}{k+1}.
\]
The upper bound on $\operatorname{sw}(W^*)$ from A.1 gives
\[
 r\operatorname{sw}(W^*)
 \leq rk|N(c^*)|
 +r\bigl(\operatorname{sw}(C')-|C'|\,|N(c^*)|\bigr).
\]
Subtracting gives
\[
 \operatorname{sw}(\overline W)-r\operatorname{sw}(W^*)
 \geq (1-r)\bigl(\operatorname{sw}(C')-|C'|\,|N(c^*)|\bigr)
      -\frac{1}{k+1}.
\]
Each member of $C'$ has at least one more supporter than $c^*$,
so the quantity in parentheses is at least $|C'|\geq1$.
Since $r\leq\nicefrac{1}{2}$ by $(\sqrt{k}-2)^2\geq0$, the
right-hand side is nonnegative.
\end{proof}

\subsection{Standard PAV and swap search}\label{app:regular_PAV}
\label{sec:standard_PAV}

Write $r=\nicefrac{2}{\sqrt{k}}-\nicefrac{2}{k}$. For size-$k$
committees, the swap argument from \Cref{prop:PAV_size_k_local_optimum}
allows us to apply the shared welfare comparison directly.

\begin{observation}\label{obs:PAV_swaps_common_threshold}
If no swap from a size-$k$ committee $W$ increases its PAV score
by more than $\varepsilon\geq0$, then $W$ is $\bound$-deletion-stable
and $(\bound+\varepsilon)$-addition-stable for some
$0\leq\bound\leq\nicefrac{n}{k}$.
\end{observation}
\begin{proof}
Let $\bound$ be the smallest marginal PAV score of a member of $W$.
The marginal scores sum to the number of represented voters, so
$k\bound\leq n$. As in the proof of \Cref{prop:PAV_size_k_local_optimum},
if adding a candidate gained more than $\bound+\varepsilon$, then
swapping it for a member with marginal score $\bound$ would gain
more than $\varepsilon$, by submodularity.
\end{proof}

\begin{observation}\label{obs:PAV_welfare_rounding}
If $\operatorname{sw}(W)\geq r\operatorname{sw}(W^*)
-\nicefrac{1}{8nk^3}$, then $\operatorname{sw}(W)\geq
r\operatorname{sw}(W^*)$.
\end{observation}
\begin{proof}
Otherwise, multiplying the welfare gap by $k$ gives
\[
 0<2\sqrt{k}\operatorname{sw}(W^*)
   -\bigl(k\operatorname{sw}(W)+2\operatorname{sw}(W^*)\bigr)
 \leq\frac{1}{8nk^2}.
\]
The two quantities in this difference have integer squares and sum
to at most $4\sqrt{k}\operatorname{sw}(W^*)\leq4nk\sqrt{k}$.
Their squares therefore differ by at most $\nicefrac{1}{2\sqrt{k}}<1$,
a contradiction.
\end{proof}

\begin{corollary}\label{cor:regular_PAV_utilitarian}
Every size-$k$ committee admitting no improving PAV swap, including
every PAV winner, has utilitarian ratio at least $r$. Bounded
local-search PAV achieves the same ratio in polynomial time with
improvement threshold $\varepsilon=\nicefrac{1}{8nk^4}$.
\end{corollary}
\begin{proof}
Since $\overline W=W$, \Cref{obs:PAV_swaps_common_threshold} and
\Cref{lem:PAV_utilitarian_comparison} give
\[
 \operatorname{sw}(W)\geq r\operatorname{sw}(W^*)-k\varepsilon.
\]
Taking $\varepsilon=0$ proves the first claim. For bounded local
search \citep{AEH+18a,KrEl24a}, the running-time argument of
\citet[Proposition~1]{AEH+18a} applies to the stated
inverse-polynomial threshold. At termination,
$k\varepsilon=\nicefrac{1}{8nk^3}$, so
\Cref{obs:PAV_welfare_rounding} gives the second claim.
\end{proof}

For $k\geq2$, this improves the bound $\nicefrac{1}{(\sqrt{k}+2)}$
of \citet[Theorem~2]{LaSk20b}. For $k=1$, PAV coincides with AV.

\section{Omitted arguments}

\subsection{A counterexample to the simple payment formula for LP-PAV}

\begin{proposition}\label{prop:LD_PAV_simple_payments_fail}
For LP-PAV with the thresholds of \Cref{cor:PAV_polynomial_guarantees},
the payment formula of \Cref{prop:PAV_simple_payments} can charge a
voter more than their budget.
\end{proposition}

\begin{proof}
Consider the profile in \Cref{tab:LD_PAV_simple_payments_fail}, with
$n=8$ and $k=5$. The thresholds are $\bound_1=\nicefrac{143}{90}$
and $\bound_2=\nicefrac{71}{45}$.

\begin{table}[htbp]
  \centering
  \begin{tabular}{c|cccc}
    Candidate    & $A_1$  & $A_2$  & $A_3$  & $A_4$--$A_8$ \\
    \hline
    $c_1$        & \cmark & \cmark & \cmark &               \\
    $c_2$--$c_3$ &        & \cmark & \cmark & \cmark        \\
    $c_4$        &        &        & \cmark & \cmark        \\
    $c_5$        & \cmark &        &        &               \\
  \end{tabular}
  \caption{A multiwinner instance with $n=8$ and $k=5$.}
  \label{tab:LD_PAV_simple_payments_fail}
\end{table}

Starting from the empty committee, LP-PAV can add $c_1,c_2,c_3,c_4$
in this order. Their marginal PAV gains are $3$, $6$,
$\nicefrac{19}{6}$, and $\nicefrac{23}{12}$, all above $\bound_1$.
In the resulting committee $W=\{c_1,c_2,c_3,c_4\}$, their marginal
PAV losses are $\nicefrac{19}{12}$, $\nicefrac{9}{4}$,
$\nicefrac{9}{4}$, and $\nicefrac{23}{12}$, all above $\bound_2$.
Adding $c_5$ gains only $\nicefrac{1}{2}$, and no swap applies since
$c_5$ has only one supporter and each winner has at least three.
Thus LP-PAV can return $W$.

Voter~1 approves only $c_1$ in $W$ and contributes $1$ to its
marginal PAV score of $\nicefrac{19}{12}$. The formula therefore
charges them $\nicefrac{12}{19}$, exceeding their budget of
$\nicefrac{5}{8}$. To see that $W$ is affordable, let voters~1--3
each pay $\nicefrac{1}{3}$ for $c_1$ and voters~4--8 each pay
$\nicefrac{1}{5}$ for each of $c_2,c_3,c_4$.
\end{proof}

\subsection{Additional Results for \Cref{tab:comparison}}
\label{app:rule_comparison}

\begin{proposition}\label{prop:comparison_counterexamples}
PAV, EP-PAV, and LP-PAV can all fail Strong EJR+.
Both EP-PAV and LP-PAV can also fail Pareto optimality.
\end{proposition}

\begin{proof}
Consider the profile in \Cref{tab:PAV_strong_EJR_counterexample}.

\begin{table}[htbp]
  \centering
  \begin{tabular}{c|cccccccc}
    Candidate & $A_1$--$A_3$ & $A_4$--$A_6$ & $A_7$ & $A_8$ & $A_9$ & $A_{10}$ & $A_{11}$ & $A_{12}$ \\
    \hline
    $c_1$--$c_2$ & \cmark & \cmark &        &        &        &        &        &        \\
    $c_3$        & \cmark &        &        &        &        &        &        &        \\
    $c_4$        &        &        & \cmark &        &        &        &        &        \\
    $c_5$        &        &        &        & \cmark &        &        &        &        \\
    $c_6$        &        &        &        &        & \cmark &        &        &        \\
    $c_7$        &        &        &        &        &        & \cmark &        &        \\
    $c_8$        &        &        &        &        &        &        & \cmark &        \\
    $c_9$        &        &        &        &        &        &        &        & \cmark \\
  \end{tabular}
  \caption{An instance with $n=12$ and $k=8$.}
  \label{tab:PAV_strong_EJR_counterexample}
\end{table}

In the full candidate set $W=C$, $c_1$ and $c_2$ each have marginal PAV score
$3\cdot\nicefrac{1}{3}+3\cdot\nicefrac{1}{2}=\nicefrac{5}{2}$,
while every other candidate has marginal score one. Since a size-eight
committee omits exactly one candidate, $W=C\setminus\{c_3\}$ is a
PAV winner.

For EP-PAV, the penalty is $\nicefrac{3}{2}$. By submodularity,
adding either $c_1$ or $c_2$ whenever it is missing gains at least
$\nicefrac{5}{2}$. Once both are selected, each other candidate
contributes only one PAV point, so the unique EP-PAV outcome is
$W'=\{c_1,c_2\}$. LP-PAV can also return $W'$, by adding $c_1$ and then $c_2$. At $W'$ both winners have marginal
score three, every addition gains one, and no omitted candidate has
a supporter set strictly containing that of either winner.

For both $W$ and $W'$, take $N'=N$ and challenger $c_3$ in the definition of Strong EJR+
(\Cref{def:strong_EJR_plus}). For voters~1--3, the sum in the definition
is $\nicefrac{1}{6}+\nicefrac{1}{6}+\nicefrac{1}{3}=\nicefrac{2}{3}$.
For every other voter it is at most $\nicefrac{1}{3}$. Thus no voter
strictly exceeds $\nicefrac{k}{n}=\nicefrac{2}{3}$, as Strong EJR+
requires. Finally, $W'\cup\{c_3\}$ is feasible and Pareto dominates
$W'$, since adding $c_3$ benefits voters~1--3 and leaves all other
voters equally satisfied.
\end{proof}

\subsection{Pruning a standard PAV outcome}

\begin{proposition}\label{prop:PAV_no_proportional_subset}
There is a PAV winner $W$ such that every subset of $W$ satisfying
EJR+ violates UQ, JUQ, and JNQ.
\end{proposition}

\begin{proof}
Consider the profile in \Cref{tab:PAV_no_proportional_subset}, with
$n=66$ and $k=18$.

\begin{table}[htbp]
  \centering
  \begin{tabular}{c|cccccccc}
    Number of voters & 13 & 13 & 9 & 9 & 9 & 9 & 2 & 2 \\
    \hline
    $c_1$--$c_6$       & \cmark & \cmark & \cmark & \cmark & \cmark & \cmark & \cmark &        \\
    $c_7$--$c_9$       & \cmark &        & \cmark &        &        &        &        &        \\
    $c_{10}$--$c_{12}$ & \cmark &        &        & \cmark &        &        &        &        \\
    $c_{13}$--$c_{15}$ &        & \cmark &        &        & \cmark &        &        &        \\
    $c_{16}$--$c_{18}$ &        & \cmark &        &        &        & \cmark &        &        \\
    $d$                & \cmark & \cmark &        &        &        &        &        &        \\
    $e$                &        &        &        &        &        &        &        & \cmark \\
  \end{tabular}
  \caption{An approval profile with $n=66$ and $k=18$.}
  \label{tab:PAV_no_proportional_subset}
\end{table}

We first show that $W=C\setminus\{d,e\}$ is a PAV winner. In the
full candidate set, each of $c_7,\ldots,c_{18}$ has marginal PAV
score $\nicefrac{13}{13}+\nicefrac{9}{9}=2$. The marginal scores
of $d$ and $e$ are also two, and those of $c_1,\ldots,c_6$ are
larger. By submodularity, removing any two candidates costs at
least four PAV points. Removing $d$ and $e$ costs exactly four,
since their supporter sets are disjoint.

Let $W'\subseteq W$ satisfy EJR+. Each of $c_1,\ldots,c_6$ has
the same 64 supporters, whose lower quota is 17. Since none of
them approves more than 12 members of $W$, omitting any of these
six candidates would violate EJR+. Moreover, $d\notin W'$ and
its 26 supporters have lower quota seven. Thus EJR+ requires
at least one of them to approve seven members of $W'$ or more.
In particular, $W'$ must contain some $c\in\{c_7,\ldots,c_{18}\}$.

We use this $c$ to witness all three violations in
\Cref{def:quota_axioms}. Candidate $c$ has 22 supporters and quota
$q_c=6$. Every $i\in N(c)$ approves $c_1,\ldots,c_6$ as well as
$c$, so $|A_i\cap W'|\geq7>\lceil q_c\rceil$. This violates UQ.

For JUQ and JNQ, take the omitted candidate $e$ and let $N'=N(e)$.
These two voters approve only $e$ and have quota
$q(N')=\nicefrac{6}{11}$. For every $i\in N'$,
\[
 |A_i\cap((W'\setminus\{c\})\cup\{e\})|
 =1=\lceil q(N')\rceil.
\]
Together with the strict upper-quota violation for $c$, this
violates JUQ. Finally, $N(c)$ and $N(e)$ are disjoint, so both
thresholds in the definition of JNQ are $\nicefrac{1}{2}$. We have
\[
 |A_i\cap W'|-q_c\geq 1>\nicefrac{1}{2}
 \quad\text{for every }i\in N(c),
 \qquad
 q(N')-|A_i\cap W'|=\nicefrac{6}{11}>\nicefrac{1}{2}
 \quad\text{for every }i\in N'.
\]
Thus the same $c$, $e$, and $N'$ also witness a JNQ violation.
\end{proof}

\end{document}